\documentstyle[12pt]{article}
\input psfig.sty
\input epsfig.sty

\begin{document} \begin{titlepage}
\rightline{\vbox{\halign{&#\hfil\cr
&SLAC-PUB-8754\cr
&January 2001\cr}}}

\begin{center} 

\bigskip\bigskip\bigskip
{\Large\bf The Higgs Mechanism in Non-commutative \\
Gauge Theories}
\footnote{Work supported by the Department of 
Energy, Contract DE-AC03-76SF00515}

\medskip

\normalsize 
{\bf  Frank J. Petriello}
\vskip .3cm
Stanford Linear Accelerator Center \\
Stanford University \\
Stanford CA 94309, USA\\
\vskip .3cm
\end{center}

\begin{abstract}

This paper investigates the non-commutative version of the Abelian Higgs model at the one loop level.  We find that the BRST invariance of the theory is maintained at this order in perturbation theory, rendering the theory one-loop renormalizable.  Upon removing the gauge field from the theory we also obtain a consistent continuum renormalization of the broken O(2) linear sigma model, contradicting results found in the literature.  The beta functions for the various couplings of the gauged U(1) theory are presented, as are the divergent contributions to every one particle irreducible (1PI) function.  We find that all physical couplings and masses are gauge independent.  A brief discussion concerning the symmetries $P$, $C$, and $T$ in this theory is also given.

\end{abstract}
\end{titlepage}

\newpage
\section{Introduction}
This paper studies the perturbative aspects of spontaneous symmetry breaking in noncommutative gauge theories.  Non-commutative field theories have been the subject of great activity recently (see, for example, references [1-34]), since being found to arise naturally in limits of string theory formulated in the presence of background gauge fields~\cite{Hull, Connes, Witten}.  Later works have treated non-commutative field theories as objects worthy of study independent of their string theory origins, a point of view adhered to in this paper.  

A hallmark of these theories is the mixing of UV and IR divergences~\cite{Seiberg1}; UV divergences in the commutative theory can become IR divergences in the noncommutative theory.  This calls into question the renormalizability of non-commutative field theories.  Several papers have explicitly shown that such theories as $\phi^4$ and U(N) gauge theories, when formulated on a non-commutative space, are one loop renormalizable~\cite{Haya,Com1,Com2,Com3,Adi,Martin}.  However, results in the literature~\cite{Campbell} have shown that non-commutativity renders impossible the continuum renormalization of the spontaneously broken linear sigma model.  They find that Goldstone's theorem is violated at the one loop level, and the Goldstone mode obtains a mass dependent upon the theory's UV cutoff.  The situtation in spontaneously broken gauge theories seems to also merit investigation, as both an interesting question and as a preface to any attempts to embed the Standard Model within a non-commutative framework.  In particular, the gauge dependence of the spontaneously broken theory should be checked, as the analog of the problem seen in~\cite{Campbell} would be a gauge dependent shift of one of the masses in the theory.  In this paper we examine the non-commutative Abelian Higgs model at the one-loop level.  We work in an arbitrary $R_{\xi}$ gauge, and show that the resulting BRST invariance of the action holds when one loop corrections are calculated by finding a counterterm set capable of removing the divergences from the 1PI functions.  We find that the physical couplings and masses are gauge-independent.  Upon taking the gauge coupling to zero, we obtain a continuum renormalization of the broken O(2) linear sigma model, contradicting the results in~\cite{Campbell}.  We show that the proper ordering of the NC generalization of $ |\phi|^4$ term in the globally symmetric theory is that consistent with the local realization of the symmetry.  We then summarize some of the properties of the theory, such as the beta functions for the various couplings and violations of the discrete symmetries $P$, $C$, and $T$ for certain types of non-commutativity.  

The paper is organized as follows.  In section 2 we briefly review the essential ideas of non-commutative (NC) field theory; for a more detailed introduction the reader is referred to the references in the bibliography.  Section 3 reviews the commutative Abelian Higgs model, concentrating upon setting up the counterterm structure and upon defining gauge independent physical parameters.  We discuss this in detail as similar definitions will be used when discussing the NC model.  In section 4 we construct the NC Abelian Higgs model, and show by explicit calculation that the theory is renormalizable.  We present our conclusions in section 5.

\section{Field Theories on NC Spaces}

The essential idea of NCQFT is a generalization of the usual d-dimensional 
space, $R^d$, associated with commuting space-time coordinates to one which 
is non-commuting, $R^d_\theta$. In such a space the conventional 
coordinates are represented by operators which no longer commute: 
\begin{equation}
[\hat X_\mu,\hat X_\nu]=i\theta_{\mu\nu},
\end{equation}
where $ \theta_{\mu \nu}$ represents a real anti-symmetric matrix.  NCQFT can be phrased in terms of conventional commuting QFT through the 
application of the Weyl-Moyal correspondence
\begin{equation}
\hat A(\hat X) \longleftrightarrow A(x)\,,   
\end{equation}
where $A$ represents a quantum field with $\hat X$ being the set of  
non-commuting coordinates and $x$ corresponding to the commuting set.  However,
in formulating  NCQFT, one must be careful to preserve orderings
in expressions such as $\hat A(\hat X)\hat B(\hat X)$.  This is accomplished
with the introduction of the Moyal product, $\hat A(\hat X)\hat B(\hat X) =
A(x) \times B(x)$, where the effect of the commutation relation is absorbed into the cross.  Introducing the Fourier transform pair
\begin{eqnarray}
\hat A(\hat X) &=& {1\over {(2\pi)^{d/2}}}\int~d\alpha e^{i\alpha \hat X}
~a(\alpha)\nonumber \\
a(\alpha) &=& {1\over {(2\pi)^{d/2}}}\int~dx e^{-i\alpha x}~A(x)\,,
\end{eqnarray}
with $x$ and $\alpha$ being real n-dimensional variables, allows us to write 
the product of two fields as 
\begin{eqnarray}
\hat A(\hat X) \hat B(\hat X) &=& {1\over {(2\pi)^{d}}}\int~d\alpha d\beta 
e^{i\alpha \hat X}~a(\alpha)e^{i\beta \hat X}~b(\beta)\nonumber \\
&=& {1\over {(2\pi)^{d}}}\int~d\alpha d\beta ~e^{i(\alpha+\beta)\hat X-
{1\over {2}}\alpha^\mu \beta^\nu[\hat X_\mu,\hat X_\nu]}a(\alpha)b(\beta)\,.
\end{eqnarray}
We thus have the correspondence 
\begin{equation}
\hat A(\hat X) \hat B(\hat X) \longleftrightarrow A(x) \times B(x)\,,   
\end{equation}
provided we identify
\begin{equation}
A(x) \times B(x)\equiv \bigg[e^{{i\over {2}}\theta_{\mu\nu}\partial_{\zeta\mu}
\partial_{\eta\nu}}A(x+\zeta)B(z+\eta)\bigg]_{\zeta=\eta=0}\,. 
\end{equation}
Note that propagators are identical on commutative and NC spaces because quadratic forms remain unchanged:
\begin{eqnarray}
\int d^{4} x \, \phi(x) \times \phi(x)=\int \phi(x) \, {\rm exp} \bigg \{\frac{i}{2}\theta_{\mu \nu} \stackrel{\leftarrow}{\partial}_{\mu} \stackrel{\rightarrow}{\partial}_{\nu} \bigg \} \, \phi(x) \nonumber \\ =\int \phi(x) \, {\rm exp} \bigg \{\frac{-i}{2}\theta_{\mu \nu} \stackrel{\rightarrow}{\partial}_{\mu} \stackrel{\rightarrow}{\partial}_{\nu} \bigg \} \, \phi(x)= \int d^{4}x \, \phi(x) \phi(x),
\end{eqnarray}
\noindent
since $\theta_{\mu \nu}$ is antisymmetric.

\section{Commutative Abelian Higgs Model}

Here we review the commutative Abelian Higgs model in some detail, as much of our construction will carry over into the non-commutative case.  The commutative Abelian Higgs model begins with the Lagrangian 
\begin{equation}
{\cal L}_{AH}= \frac{-1}{4}(F_{\mu \nu})^2 + \left|\partial_{\mu}+igA_{\mu} \right|^2 + \mu^2 \left|\phi \right|^2 -\frac{\lambda}{6} \left|\phi \right|^4,
\end{equation}
\noindent
where $F_{\mu \nu} = \partial_{\mu} A_{\nu} -\partial_{\nu} A_{\mu}$.  This Lagrangian is invariant under the gauge transformation
\begin{eqnarray} 
\phi \rightarrow e^{ie \alpha(x)} \phi, \nonumber \\
 A_{\mu} \rightarrow A_{\mu} -\partial_{\mu} \alpha.
\end{eqnarray}
The potential 
$$ V[\phi]=\mu^2 \left|\phi \right|^2 -\frac{\lambda}{6} \left|\phi \right|^4$$
has minima at 
\begin{equation}
\left|\phi_{min} \right|^2 = \nu^2 = \frac{3 \mu^2}{\lambda}.
\end{equation}
Expanding 
\begin{equation}
\phi=\nu + \frac{h}{\sqrt{2}} +\frac{i \sigma}{\sqrt{2}},
\end{equation}
we arrive at the Lagrangian
\begin{eqnarray}
{\cal L}_{AH} &=&  \frac{-1}{4}(F_{\mu \nu})^2 +\frac{1}{2}M^2 A^2 + \frac{1}{2}(\partial_{\mu} \sigma)^2 + \frac{1}{2}(\partial_{\mu} h)^2 -\frac{1}{2}m^2 h^2 \nonumber \\ & &+M A^{\mu} \partial_{\mu} \sigma  - \frac{\sqrt{2} \lambda \nu}{6} h^3 
- \frac{\sqrt{2} \lambda \nu}{6} h \sigma^2 -\frac{\lambda}{12} h^2 \sigma^2 
 - \frac{\lambda}{24}h^4 \nonumber \\ & &-\frac{\lambda}{24} \sigma^4 + \frac{g^2}{2} \sigma^2 A^2
+ \frac{g^2}{2} h^2 A^2 +g h A^{\mu} \partial_{\mu} \sigma -g \sigma A^{\mu} \partial_{\mu} h \nonumber \\ & & +\sqrt{2} g^2 \nu h A^2.
\end{eqnarray}
The Higgs field has acquired a mass $m^2 = 2 \lambda \nu^2 /3 $, while the gauge boson has acquired a mass $ M^2 = 2 g^2 \nu^2 $.  We will work in an $R_{\xi}$ gauge, so to this we add the gauge fixing and ghost Lagrangians
\begin{equation}
{\cal L}_{gf} + {\cal L}_{gh} = \frac{-1}{2 \xi} (\partial_{\mu} A^{\mu} - \xi M \sigma)^2 - {\bar c} \left( \partial^2 +\xi M^2 + \xi gMh \right) c,
\end{equation}
which cancels the $A - \sigma$ cross term.  The Feynman rules can be found in~\cite{Quinn}.  The total Lagrangian ${\cal L}_{AH} + {\cal L}_{gf} + {\cal L}_{gh}$ is invariant under the BRST transformation
\begin{eqnarray}
\delta h &=& -g \sigma c \, \Theta \nonumber \\
\delta \sigma &=& M c \, \Theta + ghc \, \Theta \nonumber \\ 
\delta A_{\mu} &=& - \left( \partial_{\mu} c \right) \Theta \nonumber \\
\delta {\bar c} &=& - \frac{1}{\xi} \left( \partial_{\mu} A^{\mu} - \xi M \sigma \right) \Theta \nonumber \\
\delta c &=& 0 \,\, .
\end{eqnarray}

To study the renormalizability of the theory we define the following counterterms relating the bare and physical quantities:
\begin{eqnarray}
 A_{B}^{\mu}=Z^{1/2}_{A} A^{\mu}, \;\;\;\; \phi_{B}=Z^{1/2}_{\phi} \phi, \;\;\;\; \mu^{2}_{B}=Z^{-1}_{\phi} Z_{\mu} \mu^2, \nonumber \\
  \, \lambda_{B} = Z^{-2}_{\phi} 
 Z_{\lambda} \lambda m_{D}^{4-d}, \;\;\;\; g_{B}= Z^{-1/2}_{A} Z_{g} g m_{D}^{2-d/2},
\end{eqnarray}
where $m_{D}$ is a constant with dimensions of mass used to account for units in dimensional regularization.  Note that if we expand 
$$\phi=\nu + \frac{h}{\sqrt{2}} +\frac{i \sigma}{\sqrt{2}}, \;\;\;\;\; \nu^2 = \frac{3 \mu^2}{\lambda}$$
as before, we will no longer be expanding around the minimum of the potential; the higgs tadpole will acquire a nonzero value.  It is convenient, though unnecessary, to introduce a new counterterm $Z_{\nu}$, expand
\begin{equation}
\phi=Z_{\nu} \nu + \frac{h}{\sqrt{2}} +\frac{i \sigma}{\sqrt{2}},
\end{equation}
and fix $Z_{\nu}$ by requiring the higgs tadpole to vanish.  We could, if desired, refrain from introducing $Z_{\nu}$, and include the Higgs tadpole in the calculation of other Green's functions.  Depending upon the gauge in which we work, $Z_{\nu}$ will be UV divergent.  Although it may seem strange to be expanding the scalar field around an infinite gauge-dependent vev, the expansion point is not a physical obervable, so no contradiction arises.  This procedure is discussed in~\cite{Collins}.
We now take the Lagrangian in eq. (8), written in terms of bare quantities, and insert the physical quantities and counterterms, while expanding the field $\phi$ as in eq. (16).  Our new Lagrangian contains two pieces: the Lagrangian of eq. (12) written in terms of the physical parameters, and the counterterm Lagrangian, which is used to subtract the divergences in the physical Green's functions.  The counterterm Lagrangian generated from the original Lagrangian plus ${\cal L}_{gf}$ is
\begin{eqnarray}
{\cal L}^{cnt}_{AH+gf}&=& \frac{1}{\sqrt{2}} \nu m^2 \left(Z_{\nu} Z_{\mu}-Z_{\nu}^3 Z_{\lambda} \right) h -\frac{1}{2} m^2 \left( \frac{3}{2}Z_{\nu}^2 Z_{\lambda} -\frac{1}{2} Z_{\mu} -1 \right) h^2  -\frac{1}{4} m^2 \left( Z_{\nu}^2 Z_{\lambda} - Z_{\mu} \right) \sigma^2 \nonumber \\ & & 
+ \frac{1}{2} \left(Z_{\phi} -1 \right) (\partial_{\mu} h)^2 + \frac{1}{2} \left(Z_{\phi}-1 \right) (\partial_{\mu} \sigma)^2 - \frac{1}{24} \left(Z_{\lambda}-1 \right) \lambda h^4 \nonumber - \frac{1}{24} \left(Z_{\lambda}-1 \right) \lambda \sigma^4 \nonumber \\ & & - \frac{1}{12} \left(Z_{\lambda}-1 \right) \lambda h^2 \sigma^2 -\frac{\sqrt{2}}{6} \left(Z_{\nu} Z_{\lambda} -1 \right) \lambda \nu h^3 -\frac{\sqrt{2}}{6} \left(Z_{\nu} Z_{\lambda} -1 \right) \lambda \nu h \sigma^2 \nonumber \\ & & - \frac{Z_A}{4} \left( F_{\mu \nu} \right)^2 + \sqrt{2} \left( Z_g Z_{\phi} Z_{\nu} -1 \right) g \nu A^{\mu} \partial_{\mu} \sigma + \left( Z_{\phi} Z_{g}^2 Z_{\nu}^2 -1 \right) g^2 \nu^2 A^2 \nonumber \\ & & + \left( Z_{\phi} Z_{\nu} Z_{g}^2 -1 \right) g^2 \nu h A^2 + \frac{1}{2} \left( Z_{\phi} Z_{g}^2 -1 \right) g^2 \sigma^2 A^2 + \frac{1}{2} \left( Z_{\phi} Z_{g}^2 -1 \right) g^2 h^2 A^2 \nonumber \\ & & + \left( Z_{\phi} Z_g -1 \right) g A^{\mu} \left[ h \partial_{\mu} \sigma - \sigma \partial_{\mu} h \right] \,\,. 
\end{eqnarray} 
This expression uses the fact that ${\cal L}_{gf}$ is already written in terms of the physical parameters and fields.  To determine the counterterms for ${\cal L}_{gh}$, we first note that the Higgs-ghost interaction is super-renormalizable, and therefore doesn't need a counterterm.  Returning to the gauge transformation of eq. (9) written in terms of the unbroken fields, and expanding as in eq. (16), we arrive at the counterterm Lagrangian
\begin{equation}
{\cal L}^{cnt}_{gh} = - \left( Z_{\nu} -1 \right) \xi M^2 {\bar c}c.
\end{equation}
The super-renormalizability of the Higgs-ghost interaction means that we do not need to introduce a wave-function renormalization constant for the ghost field.  The new Lagrangian is invariant under a ``renormalized'' BRST symmetry, which is identical to eq. (14) with $M \rightarrow Z_{\nu} M$ in the $\delta \sigma$ transformation.   

The terms listed above illustrate the subtlety involved with the renormalization of spontaneously broken theories; a limited number of counterterms are needed to subtract a large number of divergences.  The above theory is renormalizable in spite of these difficulties.  An explicit one loop calculation reveals the counterterms
\begin{eqnarray}
Z_g = 1, \,\, Z_A = 1- \frac{g^2}{24 \pi^2 \epsilon}, \;\;\; Z_{\phi} = 1+ \frac{3 g^2}{8 \pi^2 \epsilon} -\frac{\xi g^2}{8 \pi^2 \epsilon}, \;\;\; Z_{\nu} = 1 + \frac{\xi g^2}{8 \pi^2 \epsilon}, \nonumber \\
Z_{\lambda} = 1+ \frac{5 \lambda}{24 \pi^2 \epsilon} + \frac{9 g^4}{4 \pi^2 \lambda \epsilon} - \frac{\xi g^2}{4 \pi^2 \epsilon}, \;\;\; Z_{\mu} = 1+ \frac{\lambda}{12 \pi^2 \epsilon} - \frac{\xi g^2}{8 \pi^2 \epsilon},
\end{eqnarray}
where $\epsilon = 4-d$ can account for the one loop divergences in this theory, and we have used the minimal subtraction prescription.

In preparation for our discussion of the NC case, let us discuss how to obtain gauge-independent couplings and masses.  Eq. (15) gives the relations between the bare couplings and physical couplings; solving the equations for the physical couplings in terms of the bare couplings and renormalization constants, and inserting the expressions of eq. (19) for the renormalization constants, gives $ \xi $ independent expressions for the physical couplings.  The calculation of the physical couplings at various renormalization points is facilitated by finding their beta functions.  We find the following values:
\begin{eqnarray}
\beta(\lambda) &=& m_{D} \frac{\partial \lambda}{\partial m_D}= \frac{5 \lambda^2}{24 \pi^2} - \frac{3 \lambda g^2}{4 \pi^2} + \frac{9g^4}{4 \pi^2} \nonumber \\
\beta(g^2) &=& m_{D} \frac{\partial g^2}{\partial m_D}= \frac{g^4}{24 \pi^2}.
\end{eqnarray}
These quantities are in agreement with those found in~\cite{Coleman}.  We can solve these differential equations to find the relations between physical couplings at various renormalization points; for example, we find
\begin{equation}
g^2 = \frac{g_{0}^{2}}{1- \frac{g_{0}^2}{24 \pi^2} \, {\rm ln} \left( \frac{m_{D}}{m_{D0}} \right)},
\end{equation}
where $g_0$ is the coupling at the renormalization point $m_{D0}$.  Similarly for the masses, we have the following relations between bare and physical masses:
\begin{eqnarray}
m_{B}^2 &=& \frac{2}{3} \, \lambda_{B} \nu_{B}^2 = Z_{\mu} Z_{\phi}^{-1} \, m^2 \nonumber \\
M_{B}^2 &=& 2 g_{B}^2 \nu_{B}^2 = Z_{g}^{2} Z_{A}^{-1} Z_{\mu} Z_{\phi} Z_{\lambda}^{-1} \, M^2
\end{eqnarray}
Note that $ \nu_{B} = 3 \mu_{B}^2 / \lambda_{B}$; $Z_{\nu}$ just defines the shift of the expansion point, and does not enter this expression.  We can check that these lead to gauge independent definitions of the physical masses; calculation yields
\begin{eqnarray}
m^2 &=& m_{B}^2 \left[ 1 - \frac{\lambda}{12 \pi^2 \epsilon} + \frac{3 g^2}{8 \pi^2 \epsilon} \right] \nonumber \\
M^2 &=& M_{B}^2 \left[ 1 + \frac{ \lambda}{8 \pi^2 \epsilon} - \frac{5 g^2}{12 \pi^2 \epsilon} + \frac{9 g^4}{4 \pi^2 \lambda \epsilon} \right],
\end{eqnarray}
where the bare masses are infinite in order to cancel the $1/ \epsilon$ poles. The important point is the gauge independence of these results; we will find that the same definitions of the physical parameters give gauge independent results in the NC theory.

\section{Non-commutative Abelian Higgs Model}

\subsection{Setup of NC Symmetry Breaking in U(1)}

We now examine the non-commutative extension of the Abelian Higgs model, following the procedure introduced in the previous section.  Non-commutative U(1) gauge theory coupled to a complex scalar field is defined by the Lagrangian
\begin{equation}
{\cal L}_{AH} = \frac{-1}{4} F_{\mu \nu} \times F^{\mu \nu} + D_{\mu} \phi \times (D^{\mu} \phi)^{*} + \mu^2 \left| \phi \right|^2 -\frac{\lambda}{6} \phi^{*} \times \phi \times \phi^{*} \times \phi,
\end{equation}
where $ D_{\mu} \phi = \partial_{\mu} \phi + i g A_{\mu} \times \phi \,$ and $F_{\mu \nu} = \partial_{\mu}A_{\nu}-\partial_{\nu}A_{\mu} +ig \left( A_{\mu} \times A_{\nu} - A_{\nu} \times A_{\mu} \right)$.  With $U(x) = e^{i g \alpha(x)}$, the action is invariant under the gauge transformation
\begin{equation}
\phi \rightarrow U \times \phi, \,\, \phi^{*} \rightarrow \phi^{*} \times U^{-1}, \,\, A_{\mu} \rightarrow U \times A_{\mu} \times U^{-1} +\frac{i}{g} (\partial_{\mu} U) \times U^{-1} .
\end{equation}
Note that of the two possible noncommutative generalizations of $\left| \phi \right|^4$, $\phi^{*} \times \phi \times \phi^{*} \times \phi$ and $\phi^{*} \times \phi^{*} \times \phi \times \phi$, only the first is consistent with local gauge invariance as defined by the transformation of eq. (16) We examined the theory with the potential $f \, \phi^* \times \phi \times \phi^* \times \phi + (1-f) \, \phi^* \times \phi^* \times \phi \times \phi$ and found that the theory is one-loop renormalizable only if $f=1$ (See also the discussion following eq. (28)).  The minimum of the potential $V[\phi]$ is the same as in the commutative theory, and as quadratic forms are unchanged by the noncommutativity, the Higgs particle and gauge boson acquire the same masses as in the commutative theory.  Expanding around the minimum $\nu$, we arrive at the Lagrangian
\begin{eqnarray} 
{\cal L}_{AH} &=& \frac{-1}{4} F_{\mu \nu} \times F^{\mu \nu} +\frac{1}{2} (\partial_{\mu} h)^2 + \frac{1}{2} (\partial_{\mu} \sigma)^2 -M \sigma \partial_{\mu} A^{\mu} +\frac{1}{2} M^2 A^2  \nonumber \\ & & - \frac{1}{2} m^2 h^2 - \frac{\sqrt{2} \nu}{6} h \times h \times h - \frac{\sqrt{2} \nu}{6} h \times \sigma \times \sigma \nonumber \\ & & - \frac{\lambda}{24} h \times h \times h \times h - \frac{\lambda}{24} \sigma \times \sigma \times \sigma \times \sigma - \frac{\lambda}{6} h \times h \times \sigma \times \sigma \nonumber \\ & & + \frac{\lambda}{12} h \times \sigma \times h \times \sigma + g M h \times A_{\mu} \times A^{\mu} + \frac{1}{2} g^2 h \times h \times A_{\mu} \times A^{\mu} \nonumber \\ & & + \frac{1}{2} g^2 \sigma \times \sigma \times A_{\mu} \times A^{\mu} + \frac{i}{2} g^2 [h, \sigma ] \times A_{\mu} \times A^{\mu} + \frac{i}{2} g A_{\mu} \times [h, \partial^{\mu} h] \nonumber \\ & & + \frac{i}{2} g A_{\mu} \times [ \sigma, \partial^{\mu} \sigma ] + \frac{1}{2} g A_{\mu} \times \left\{ h, \partial^{\mu} \sigma \right\} 
- \frac{1}{2} g A_{\mu} \times \left\{ \sigma, \partial^{\mu} h \right\},
\end{eqnarray} 
where we have used the notation $[x,y] = x \times y - y \times x$ and $ \left\{ x,y \right\} = x \times y + y \times x$.  To this we add the gauge-fixing and ghost Lagrangians
\begin{eqnarray}
 {\cal L}_{gf} + {\cal L}_{gh} &=& - \frac{1}{2 \xi} (\partial_{\mu} A^{\mu} - \xi M \sigma)^2 - {\bar c} \left( \partial^2 +\xi M^2 \right) c - \frac{\xi g M}{2} {\bar c} \left\{c,h \right\} \nonumber \\ & & - \frac{i \xi g M}{2} {\bar c} \left[c, \sigma \right] + ig {\bar c} \partial^{\mu} \left[ c,A_{\mu} \right]\,\, ;
\end{eqnarray}
to obtain the ghost Lagrangian we simply insert the BRST transformation of eq. (28) into the gauge- fixing condition 
$$ F[A, \sigma]=\partial_{\mu}A^{\mu} - \xi M \sigma \,\, .$$
The full Lagrangian ${\cal L}_{AH} + {\cal L}_{gf} + {\cal L}_{gh}$ is found to be invariant under the BRST transformation
\begin{eqnarray}
\delta h &=& - \frac{g}{2} \left\{ c, \sigma \right\} \Theta + \frac{ig}{2} \left[ c,h \right] \Theta \nonumber \\
\delta \sigma &=& Mc \, \Theta + \frac{g}{2} \left\{ c,h \right\} \Theta + \frac{ig}{2} \left[ c, \sigma \right] \Theta \nonumber \\
\delta A_{\mu} &=& - \left( \partial_{\mu} c \right) \Theta + ig \left[ c,A_{\mu} \right] \Theta \nonumber \\
\delta {\bar c} &=& - \frac{1}{\xi} \left( \partial_{\mu} A^{\mu} - \xi M \sigma \right) \Theta \nonumber \\
\delta c &=& -ig c \times c \, \Theta.
\end{eqnarray}
Note that upon replacing $c \, \Theta \rightarrow \alpha$ in the transformation laws for $h$, $\sigma$, and $A_{\mu}$, the above reduce to the infinitesimal transformations found in eq. (25).  Let us carefully show how to obtain the transformations of $h$ and $ \sigma$.  The infinitesimal form of eq. (25) is 
\begin{eqnarray}
\phi & \rightarrow & \phi^{'} = \phi + ig \, \alpha \times \phi \nonumber \\
\phi^{*} & \rightarrow & \phi^{* \, '} = \phi^{*} -ig \, \phi^{*} \times \alpha,
\end{eqnarray}
where the prime indicates the transformed field.  Written in terms of $h$ and $\sigma $, these become
\begin{eqnarray}
 \nu + \frac{1}{ \sqrt{2}} \left( h^{'} + i \sigma^{'} \right) &=& \nu + \frac{1}{\sqrt{2}} \left( h + i \sigma \right) +ig \nu \alpha + \frac{ig}{\sqrt{2}}\, \alpha \times \left(h + i \sigma \right) \nonumber \\
 \nu + \frac{1}{\sqrt{2}} \left( h^{'} - i \sigma^{'} \right) &=& \nu + \frac{1}{\sqrt{2}} \left( h - i \sigma \right) -ig \nu \alpha - \frac{ig}{\sqrt{2}} \left(h - i \sigma \right) \times \alpha.
\end{eqnarray}
Adding and subtracting these give the transformations of $h$ and $ \sigma$, respectively.

The Feynman rules for the NC Abelian Higgs model derived from the Lagrangians of eqs. (26-27) are presented in the appendix, where we introduce the notation $p \wedge q = p^{\mu} q^{\nu} \Theta_{\mu \nu} \, /2$.  Note the presence of new interactions, such as the $A-2h$ and $2A-h- \sigma$ vertices, since the terms containing commutators vanish in the commutative limit.  These arise from the violation of charge conjugation symmetry, as will be shown in the next several paragraphs.

Let us now briefly discuss the role of the discrete symmetries $P$, $C$, and $T$ in this theory; the presentation will very closely follow that in~\cite{Sheikh3}.  The transformations of the fields under the various symmetries can be derived from the requirement that the commutative Lagrangian of eq. (12) be invariant under any of the symmetries; the results are
\begin{eqnarray}
P \, h \, P^{-1} = h & & C \, h \, C^{-1} = h \;\;\;\;\;\;\;\; T \, h \, T^{-1} = h \nonumber \\ P \, \sigma \, P^{-1} = - \sigma & &  C \, \sigma \, C^{-1} = - \sigma \;\;\;\;\;\; T \, \sigma \, T^{-1} = - \sigma \nonumber \\ P \, A^{\mu} \, P^{-1} = A_{\mu} & &  C \, A_{\mu} \, C^{-1} = -A_{\mu} \;\;\;\;\;\; T \, A^{\mu} \, T^{-1} = A_{\mu}. 
\end{eqnarray}  
We must now determine whether the interactions introduced by the non-commutativity respect these symmetries.  Although the matrix $ \theta_{\mu \nu}$ is just a set of real parameters and is not affected by any of the transformations, it will be useful to indicate the transformations of the various $\theta$ parameters that would lead to an invariant NC Lagrangian, as in~\cite{Sheikh3}.

{\it Parity}: The net effect of parity on the NC Lagrangian is to change the Moyal product as follows:
\begin{equation}
{\rm exp} \left(\frac{i \theta_{0i}}{2} \stackrel{\leftarrow}{\partial}_0 \stackrel{\rightarrow}{\partial}_i + \frac{i \theta_{ij}}{2} \stackrel{\leftarrow}{\partial}_i \stackrel{\rightarrow}{\partial}_j \right) \rightarrow {\rm exp} \left( -\frac{i \theta_{0i}}{2} \stackrel{\leftarrow}{\partial}_0 \stackrel{\rightarrow}{\partial}_i + \frac{i \theta_{ij}}{2} \stackrel{\leftarrow}{\partial}_i \stackrel{\rightarrow}{\partial}_j \right). 
\end{equation}
Hence, the NC theory is $P$ invariant only if $\theta_{0i}=0$.  The theory would be parity invariant if we also took $\theta_{0i} \rightarrow - \theta_{0i}$.

{\it Charge Conjugation}: Charge conjugation leaves all but four terms of the Lagrangian invariant: the terms in eq. (26) containing commutators and the term leading to the triple gauge photon vertex, which also contains a single commutator.  These acquire a minus sign under $C$.  Charge conjugation invariance is therefore violated for any non-zero value of $\theta$; the $A-2h$, $A-2 \sigma$, $3A$, and $2A-h-\sigma$ vertices arising from the commutator terms in the Lagrangian explicitly show this violation.  We could maintin $C$ invariance by requiring $C \, \theta_{\mu \nu} \, C^{-1} = - \theta_{\mu \nu}$.
 
{\it Time Reversal}: The net effect of the time reversal invariance on the NC Lagrangian is to change the Moyal product,
\begin{equation}
{\rm exp} \left(\frac{i \theta_{0i}}{2} \stackrel{\leftarrow}{\partial}_0 \stackrel{\rightarrow}{\partial}_i + \frac{i \theta_{ij}}{2} \stackrel{\leftarrow}{\partial}_i \stackrel{\rightarrow}{\partial}_j \right) \rightarrow {\rm exp} \left( \frac{i \theta_{0i}}{2} \stackrel{\leftarrow}{\partial}_0 \stackrel{\rightarrow}{\partial}_i - \frac{i \theta_{ij}}{2} \stackrel{\leftarrow}{\partial}_i \stackrel{\rightarrow}{\partial}_j \right). 
\end{equation}
The theory is time reversal invariant only if $ \theta_{ij}=0$. $T$ invariance could be maintained by requiring $T \, \theta_{ij} \, T^{-1} = - \theta_{ij}$.

We can see from these transformations that the theory is $CPT$ invariant for all $\theta_{\mu \nu}$, and $CP$ invariant only if $ \theta_{ij} =0$.  This leads to the question of whether theories with $\theta_{ij} \neq 0$ might be used as models of CP violation in particle physics, a point raised in~\cite{Sheikh3}.  We make no attempt to address this question, as it would require the non-commutative extension of the electroweak theory, but do provide a rough estimate of the size of such effects, obtained by considering the $C$ violating process $A \rightarrow hh$.  A short calculation reveals a partial width of the form
\begin{equation}
\Gamma \sim M \, \frac{M^4}{\Lambda^4},
\end{equation}
where $ \Lambda$ is the energy scale associated with the non-commutativity.  We will not attempt to study further the phenomonology of NC theories; preliminary discussions can be found in~\cite{Sheikh4, Sheikh5,Me,Brazil,Greene,Russian}

To study the renormalization of the theory we introduce the same wave-function and coupling constant rescalings as in eq. (15).  As the ghost-gauge boson vertex contains factors of momenta, we must also introduce the ghost wave-function renormalization
\begin{equation}
c_{B} = Z_{c} c .
\end{equation}
The counterterm Lagrangian ${\cal L}^{cnt}_{AH}$ obtained from eq. (26) is the same as that of eq. (17) with multiplication replaced by the Moyal product and the anti-commutators of eq. (26) accounted for appropriately.  In addition, there are new counterterms for the three and four-point gauge boson vertices, and for the new interactions represented by commutators of eq. (26).  As the counterterm Lagrangian is rather long we will not write it explicitly; it is apparent how to obtain it, and the counterterms for each vertex are presented in section 4.2.  The ghost counterterm Lagrangian, ${\cal L}^{cnt}_{gh}$, is 
\begin{eqnarray}
{\cal L}^{cnt}_{gh} &=& - \left( Z_c -1 \right) \partial^2 c - \xi \left( Z_c Z_g Z_{\nu} -1 \right) M^2 {\bar c} c + i \left( Z_c Z_g -1 \right) g {\bar c} \, \partial^{\mu} \left[ c,A_{\mu} \right] \nonumber \\ & & - \frac{ \xi gM}{2} \left(Z_c Z_g -1 \right) {\bar c} \left\{c,h \right\}  - \frac{i \xi gM}{2} \left(Z_c Z_g -1 \right) {\bar c} \left\{c, \sigma \right\}.
\end{eqnarray}
The entire Lagrangian, ${\cal L}_{gh}+{\cal L}^{cnt}_{gh} +{\cal L}_{AH} + {\cal L}^{cnt}_{AH} + {\cal L}_{gf}$, is now invariant under the renormalized BRST transformation
\begin{eqnarray}
\delta_R h &=& - \frac{Z_c Z_g g}{2} \left\{ c, \sigma \right\} \Theta + \frac{i Z_c Z_g g}{2} \left[ c,h \right] \Theta \nonumber \\
\delta_R \sigma &=& Z_c Z_g Z_{\nu} Mc \, \Theta + \frac{Z_c Z_g g}{2} \left\{ c,h \right\} \Theta + \frac{iZ_c Z_g g}{2} \left[ c, \sigma \right] \Theta \nonumber \\
\delta_R A_{\mu} &=& - Z_c \left( \partial_{\mu} c \right) \Theta + iZ_c Z_g g \left[ c,A_{\mu} \right] \Theta \nonumber \\
\delta_R {\bar c} &=& - \frac{1}{\xi} \left( \partial_{\mu} A^{\mu} - \xi M \sigma \right) \Theta \nonumber \\
\delta_R c &=& -iZ_c Z_g g c \times c \, \Theta,
\end{eqnarray}
which arises from the BRST invariance of the Lagrangian written in terms of bare fields.  To demonstrate that this BRST invariance holds at the one-loop level, we must find a set of renormalization constants that can simultaneously remove the divergences from every 1PI function; we do this below.

\subsection{Calculation of NC divergences}

Below we present the somewhat lengthy list of the UV divergent parts of the 1PI functions; the diagrams contributing to each are summarized in the appendix.  A quick glance at the appendix will convince the reader of our wisdom in not including the intermediate stages in these calculations.  We use dimensional regularization with $d=4- \epsilon $, and the MS prescription.  As we are interested only in the UV divergences any loop integral containing ${\rm exp}(ip \wedge k)$, where $k$ is the loop momentum, will be ignored, as it is damped for large $k$ when a convergence factor is included.  We include the counterterms that must account for each divergence.  Only the distinct vertices are listed; for example, the $4-h$ and $4- \sigma$ UV divergences, $\Gamma^{4h}$ and $\Gamma^{4 \sigma}$, are identical, and only $\Gamma^{4h}$ is given.  Similarly, the following pairs of vertices are identical: the $2A-2 \sigma$ and the $2A-2h$, and the $2h-A$ and the $2 \sigma -A$.  We have also checked that 1PI functions for which no counterterms appear, such as $\Gamma^{hA}$, are UV finite.  Note that the momenta and indices appearing in the expressions below are the same as those appearing in the relevant Feynman rules in the appendix.

\begin{eqnarray}
\Gamma^{h} &=& \frac{i \lambda \nu m^2}{8 \sqrt{2} \pi^2 \epsilon} + \frac{i \xi \lambda \nu M^2}{24 \sqrt{2} \pi^2 \epsilon} + \frac{3igM^3}{8 \pi^2 \epsilon} + \frac{i \nu m^2}{\sqrt{2}} \left[ Z_{\nu} Z_{\mu} - Z_{\nu}^3 Z_{\lambda} \right] \nonumber \\
\Gamma^{2h} &=& \frac{-3ig^2 p^2 }{8 \pi^2 \epsilon } + \frac{i \xi g^2 p^2 }{8 \pi^2 \epsilon } + \frac{7i \lambda m^2 }{48 \pi^2 \epsilon } + \frac{3ig^2 M^2 }{4 \pi^2 \epsilon } + \frac{i \xi \lambda M^2 }{24 \pi^2 \epsilon }  - \frac{i \xi g^2 m^2 }{16 \pi^2 \epsilon } \nonumber \\ & &  - \, im^2 \left[ \frac{3}{2} Z_{\nu }^2 Z_{\lambda } - \frac{1}{2} Z_{\mu } -1 \right] + \, ip^2 \left(Z_{\phi} -1 \right) \nonumber \\
\Gamma^{3h} &=& \bigg( {\rm cos}(p_1 \wedge p_2) \, + \, {\rm cos}(p_1 \wedge p_3) \, + \, {\rm cos}(p_3 \wedge p_2) \bigg) \,\, \bigg[ \frac{i \nu \lambda^2}{18 \sqrt{2} \pi^2 \epsilon}  + \frac{3ig^3 M}{8 \pi^2 \epsilon} \nonumber \\ & & - \frac{i \sqrt{2} \xi \nu \lambda g^2}{24 \pi^2 \epsilon} - \frac{i \sqrt{2} \nu \lambda}{3} \left( Z_{\nu} Z_{\lambda} -1 \right) \bigg] \nonumber \\
\Gamma^{4h} &=& \bigg( {\rm cos}(p_1 \wedge p_2) \, {\rm cos}(p_3 \wedge p_4) \, + \, {\rm cos}(p_1 \wedge p_3) \, {\rm cos}(p_2 \wedge p_4) \, + \, {\rm cos}(p_1 \wedge p_4) \, {\rm cos}(p_3 \wedge p_2) \bigg) \nonumber \\ & & \times \,\, \bigg[ \frac{i \lambda^2}{36 \pi^2 \epsilon} + \frac{3ig^4}{8 \pi^2 \epsilon} - \frac{i \xi \lambda g^2}{12 \pi^2 \epsilon} - \frac{i \lambda (Z_{\lambda}-1)}{3} \bigg] \nonumber \\
\Gamma^{2 \sigma} &=& \frac{-3ig^2 p^2 }{8 \pi^2 \epsilon } + \frac{i \xi g^2 p^2 }{8 \pi^2 \epsilon } + \frac{i \lambda m^2 }{16 \pi^2 \epsilon } + \frac{3ig^2 M^2 }{8 \pi^2 \epsilon } + \frac{i \xi g^2 m^2 }{16 \pi^2 \epsilon }  - \, \frac{im^2}{2} \left[ Z_{\nu }^2 Z_{\lambda }  - Z_{\mu } \right]  \nonumber \\ & & + \, ip^2 \left(Z_{\phi} -1 \right) \nonumber \\
\Gamma^{h -2 \sigma} &=& {\rm cos}(p_1 \wedge p_2) \left[ \frac{i \lambda^2 \nu \sqrt{2}}{36 \pi^2 \epsilon} + \frac{3ig^3 M}{8 \pi^2 \epsilon} - \frac{i \xi \lambda g \nu \sqrt{2}}{24 \pi^2 \epsilon} - \frac{i \lambda \nu \sqrt{2}}{3} \left( Z_{\lambda}Z_{\nu}-1 \right) \right] \nonumber \\
\Gamma^{2h - 2 \sigma} &=& \left( 2 \, {\rm cos}(p_1 \wedge p_2) \, {\rm cos}(p_3 \wedge p_4) - {\rm cos}(p_1 \wedge p_3 + p_2 \wedge p_4) \right) \bigg[ \frac{i \lambda^2}{36 \pi^2 \epsilon} \nonumber \\ & & + \frac{3ig^4}{8 \pi^2 \epsilon} - \frac{i \xi \lambda g^2}{12 \pi^2 \epsilon}  - \frac{i \lambda}{3} \left(Z_{\lambda} -1 \right) \bigg] \nonumber \\
\Gamma^{2h -A} &=& \left[ p_1 - p_2 \right]_{\mu} {\rm sin}(p_1 \wedge p_2)  \left[ - \frac{3g^3}{16 \pi^2 \epsilon} \left(1-\xi \right) + g \left( Z_g Z_{\phi} -1 \right) \right] \nonumber \\
\Gamma^{2h-2A} &=& {\rm cos}(p_1 \wedge p_2) \, {\rm cos}(p_3 \wedge p_4) \, g_{\mu \nu} \, \left[ \frac{i \xi g^4}{2 \pi^2 \epsilon} + 2ig^2 \left( Z_{\phi} Z_{g}^2 -1 \right) \right] \nonumber \\
\Gamma^{h- \sigma -2A} &=& {\rm cos}(p_1 \wedge p_2) \, {\rm sin}(p_3 \wedge p_4) \, g_{\mu \nu} \, \left[ \frac{i \xi g^4}{2 \pi^2 \epsilon} + 2ig^2 \left( Z_{\phi} Z_{g}^2 -1 \right) \right] \nonumber \\
\Gamma^{h- \sigma -A} &=& \left[ p_1 - p_2 \right]_{\mu} {\rm cos}(p_1 \wedge p_2)  \left[ - \frac{3g^3}{16 \pi^2 \epsilon} \left(1-\xi \right) + g \left( Z_g Z_{\phi} -1 \right) \right] \nonumber \\
\Gamma^{h -2A} &=& {\rm cos}(p_1 \wedge p_2) \, g_{\mu \nu} \, \left[ \frac{ i \xi g^3 M}{4 \pi^2 \epsilon} + 2igM \left( Z_{\phi} Z_g^2 Z_{\nu} -1 \right) \right] \nonumber \\
\Gamma^{ \sigma -A} &=& p_{\mu} \, \left[ - \frac{3 g^2 M}{16 \pi^2 \epsilon} + \frac{\xi g^2 M}{16 \pi^2 \epsilon} + M \left( Z_{\phi} Z_g Z_{\nu} -1 \right) \right] \nonumber \\
\Gamma^{c - {\bar c}} &=& - \frac{3ig^2 p^2}{16 \pi^2 \epsilon} + \frac{i \xi g^2 p^2}{16 \pi^2 \epsilon} + ip^2 \left( Z_c -1 \right) - i \left(Z_c Z_g Z_{\nu} -1 \right) \xi M^2 \nonumber \\
\Gamma^{c - {\bar c} -A} &=& p_{2}^{\mu} \, {\rm sin}(p_1 \wedge p_2) \left[ \frac{ \xi g^3}{4 \pi^2 \epsilon} + 2g \left( Z_c Z_g -1 \right) \right] \nonumber \\
\Gamma^{c - {\bar c} -h} &=& {\rm cos}(p_1 \wedge p_2) \left[ - \frac{i \xi^2 g^3 M}{8 \pi^2 \epsilon} -i \xi gM \left(Z_c Z_g -1 \right) \right] \nonumber \\
\Gamma^{c - {\bar c} -\sigma} &=& {\rm sin}(p_1 \wedge p_2) \left[ - \frac{i \xi^2 g^3 M}{8 \pi^2 \epsilon} -i \xi gM \left(Z_c Z_g -1 \right) \right] \nonumber \\
\Gamma^{2A} &=& \left( g_{\mu \nu} p^2 - p_{\mu} p_{\nu} \right) \left[ \frac{ig^2}{2 \pi^2 \epsilon} - \frac{i \xi g^2}{8 \pi^2 \epsilon} \right] + i g_{\mu \nu} \, \left( Z_{\phi} Z_{g}^2 Z_{\nu}^2 -1 \right) M^2 \nonumber \\ & & -i \left(Z_A -1 \right) \left[ g_{\mu \nu} p^2 - p_{\mu} p_{\nu} \right] \nonumber \\
\Gamma^{3A} &=& {\rm sin}(p_1 \wedge p_2) \, \bigg\{ \left( p_1 - p_2 \right)_{\rho} g_{\mu \nu} + \left( p_2 - p_3 \right)_{\mu} g_{ \nu \rho } + \left( p_3 - p_1 \right)_{\nu} g_{ \mu \rho } \bigg\} \nonumber \\ & & \times \left[ - \frac{5g^3}{8 \pi^2 \epsilon} + \frac{3 \xi g^3}{8 \pi^2 \epsilon} + 2g \left( Z_A Z_g -1 \right) \right] \nonumber \\
\Gamma^{4A} &=& \bigg\{ \, {\rm sin}(p_1 \wedge p_2) \, {\rm sin}(p_3 \wedge p_4) \, \left[ g_{\mu \rho}g_{\nu \sigma} - g_{\mu \sigma}g_{\nu \rho} \right] +  {\rm sin}(p_3 \wedge p_1) \, {\rm sin}(p_2 \wedge p_4) \nonumber \\ & & \times \, \left[ g_{\mu \nu}g_{\rho \sigma} - g_{\mu \rho}g_{\nu \sigma} \right]  +  {\rm sin}(p_1 \wedge p_4) \, {\rm sin}(p_2 \wedge p_3) \, \left[ g_{\mu \sigma}g_{\nu \rho} - g_{\mu \nu}g_{\rho \sigma} \right] \bigg\} \nonumber \\ & & \times \left[ \frac{ig^4}{2 \pi^2 \epsilon} - \frac{i \xi g^4}{ \pi^2 \epsilon} - 4ig^2 \left( Z_A Z_{g}^2 -1 \right) \right]
\end{eqnarray}
We remind the reader of the relations $ m^2 = 2 \lambda \nu^2 /3$, $M^2 = 2 g^2 \nu^2$, and $ M^2 / m^2 = 3 g^2 / \lambda$, which are used in showing that the renormalization constants listed below can remove these divergences.  We would also like to point out that our expressions for individual diagrams agree with~\cite{Martin} when applicable.  An interesting feature of these results is that individual diagrams were not necessarily proportional to the momentum dependent phase present in the vertices.  This is particularly striking in the $4A$ vertex; it contains a very non-trivial Lorentz index and phase factor structure, and receives contributions from a very large number of diagrams, none of which are proportional to the necessary factor.  This point was also emphasized in~\cite{Martin}, who found the same behavior in pure NC U(N) gauge theories in arbitrary Lorentz gauges.  

There are a very limited number of renormalization constants that must account for a large number of divergences; we find, however, that the following set suffices:
\begin{eqnarray}
Z_{\lambda} &=& 1 + \frac{\lambda}{12 \pi^2 \epsilon} + \frac{9g^4}{8 \pi^2 \lambda \epsilon} - \frac{\xi g^2}{4 \pi^2 \epsilon} \nonumber \\
Z_{\mu} &=& 1 - \frac{\lambda}{24 \pi^2 \epsilon} - \frac{9 g^4}{8 \pi^2 \lambda \epsilon} - \frac{\xi g^2}{8 \pi^2 \epsilon} \nonumber \\
Z_{\phi} &=& 1 + \frac{3 g^2}{8 \pi^2 \epsilon} - \frac{\xi g^2}{8 \pi^2 \epsilon} \nonumber \\
Z_{\nu} &=& 1 + \frac{\xi g^2}{8 \pi^2 \epsilon} \nonumber \\
Z_A &=& 1 + \frac{g^2}{2 \pi^2 \epsilon} - \frac{\xi g^2}{8 \pi^2 \epsilon} \nonumber \\
Z_g &=& 1 - \frac{3 g^2}{16 \pi^2 \epsilon} - \frac{\xi g^2}{16 \pi^2 \epsilon} \nonumber \\
Z_c &=& 1+ \frac{3 g^2}{16 \pi^2 \epsilon} - \frac{\xi g^2}{16 \pi^2 \epsilon}.
\end{eqnarray}
The relations between the divergent pieces of the 1PI functions established by the BRST symmetry of eq. (28) account for the renormalizability.  The same definitions as in the commutative theory give the following beta functions and physical masses:
\begin{eqnarray}
\beta(\lambda) &=& m_{D} \frac{\partial \lambda}{\partial m_D}= \frac{\lambda^2}{12 \pi^2} - \frac{3 \lambda g^2}{4 \pi^2} + \frac{9g^4}{8 \pi^2} \nonumber \\
\beta(g^2) &=& m_{D} \frac{\partial g^2}{\partial m_D}= \frac{-7g^4}{8 \pi^2} \nonumber \\ 
m^2 &=& m_{B}^2 \left[ 1 + \frac{\lambda}{24 \pi^2 \epsilon} + \frac{3 g^2}{8 \pi^2 \epsilon} + \frac{9 g^4}{8 \pi^2 \lambda \epsilon} \right] \nonumber \\
M^2 &=& M_{B}^2 \left[ 1 + \frac{\lambda}{8 \pi^2 \epsilon} + \frac{g^2}{2 \pi^2 \epsilon} + \frac{9 g^4}{4 \pi^2 \lambda \epsilon} \right].
\end{eqnarray}
The U(1) coupling remains asymptotically free, as in the free NC U(1) theory.  As in the commutative case, we are able to define gauge independent couplings and masses.

We can also use these results to discuss spontaneously broken global symmetries in NC field theories.  Upon removing the gauge field and gauge-scalar couplings from our Lagrangian, and making the gauge transformation global, we are left with the broken O(2) linear sigma model.  The remaining renormalization constants are $Z_{\phi}$, $Z_{\mu}$, and $Z_{\lambda}$ (with $ \xi = g = 0$); the familiar wave-function, mass, and coupling constant renormalizations that are found in textbooks~\cite{Peskin}.  Our results show that the continuum renormalization of this model is possible, contradicting results found in~\cite{Campbell}.  While in the case of global symmetries both NC generalizations of $ | \phi |^4$ discussed below eq. (25) are consistent with the symmetry, our result indicates that the proper ordering is the one also consistent with a local realization of the symmetry.  This is the only choice that leads to a one-loop renormalizable theory.  We imagine that such considerations in choosing non-commutative extensions of commutative interactions hold generally.

\section{Conclusion}

We have found that the relations between counterterms which are necessary to renormalize spontaneously broken U(1) gauge theory occur in the noncommutative version of the theory; the BRST symmetry of the Lagrangian holds at the one-loop level.  Upon taking the gauge field couplings to zero we obtain a consistent continuum renormalization of the broken O(2) linear sigma model, in disagreement with~\cite{Campbell} (see also~\cite{Campbell2}).  In the O(2) linear sigma model, both NC generalizations of the $ \ | \phi |^4 $ preserves the symmetry; however, renormalization requires us to pick the one also consistent with the local symmetry.  We are not familiar with any discussions in the literature regarding how to choose NC extensions of commutative actions when some symmetry does not dictate a choice.  However, we believe that the problem of ordering ambiguities arising from NC extensions of global symmetries can be solved by demanding that the local symmetry also hold.  Note that this wouldn't have dictated a choice of scalar potential in~\cite{Com3}, as either $ | \phi |^4 $ generalization is consistent when working in the adjoint representation.

We have only discussed a single simple case in this paper, that of a U(1) NC gauge theory coupled to a complex scalar field in the fundamental representation.  Important generalizations are to consider different scalar representations, fermion contributions, and arbitrary U(N) groups.  While we have nothing to say about the first two, we believe that the generalization to U(N) will be successful.  The remarkable interplay between diagrams required to renormalize the $4A$ vertex, seen here in the scalar sector and in the gauge sector for general U(N) in~\cite{Martin} seems to indicate the consistency of these models.

The work in this paper should be regarded as a ``proof of principle'' that spontaneously broken NC gauge theories are consistent.  The non-commutativity of space-time at small scales is an exciting possible modification of fundamental physics.  Our result provides a step towards a NC version of the Standard Model; the work in~\cite{Sheikh4,Sheikh5,Me,Brazil,Greene,Russian}, and the presence of tree level CP violation noted here and in~\cite{Sheikh3}, indicates that it might lead to interesting physics.  Hopefully this paper will direct interest towards exploring this route.

\medskip
\noindent{\Large\bf Acknowledgements}

It is a pleasure to acknowledge the following people: JoAnne Hewett for reading the manuscript and offering many helpful suggestions, Michael Peskin for reading an earlier version and finding a critical error, Tom Rizzo for numerous discussions and insightful comments, and Darius Sadri for discussions concerning discrete symmetries in particle physics.  This work was supported in part by a NSF graduate research fellowship.

\newpage

\noindent{\Large\bf Appendix A: Feynman rules}
\noindent
\begin{figure}[htbp]
\centerline{
\psfig{figure=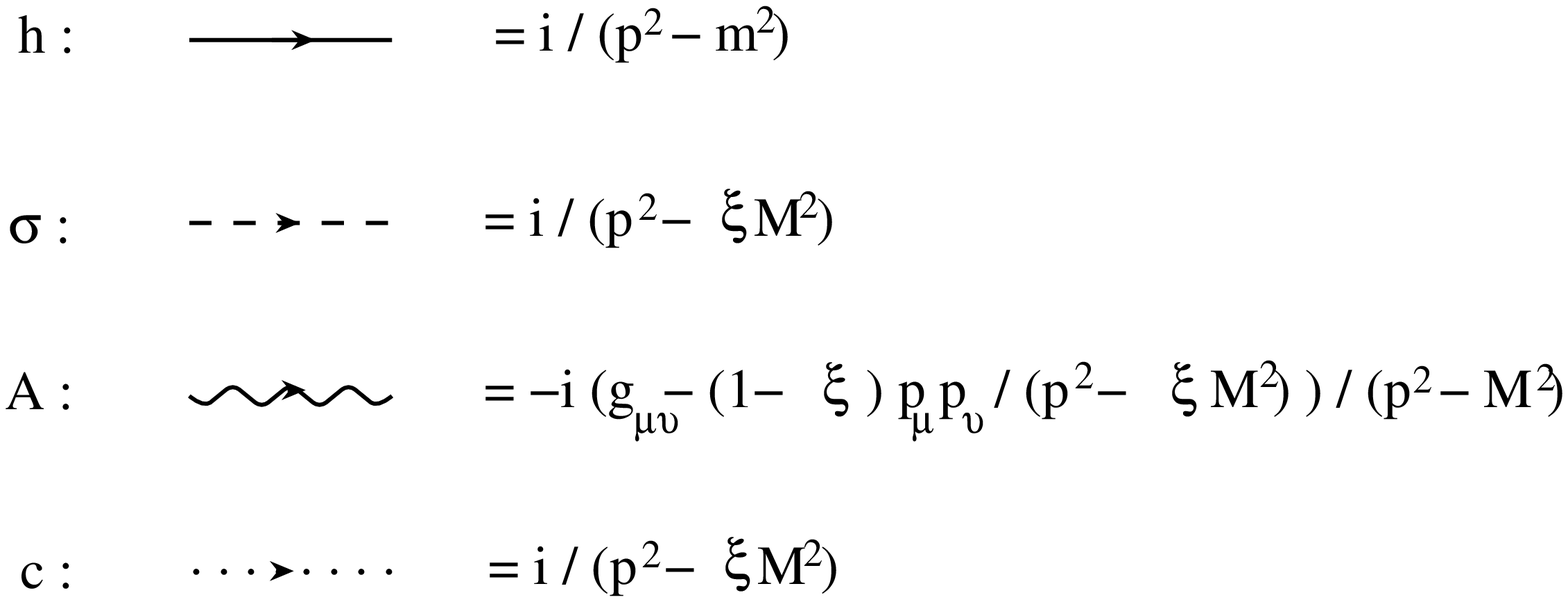,height=10.0cm,width=15.5cm,angle=0}}
\vspace*{-2.2cm}
\centerline{
\psfig{figure=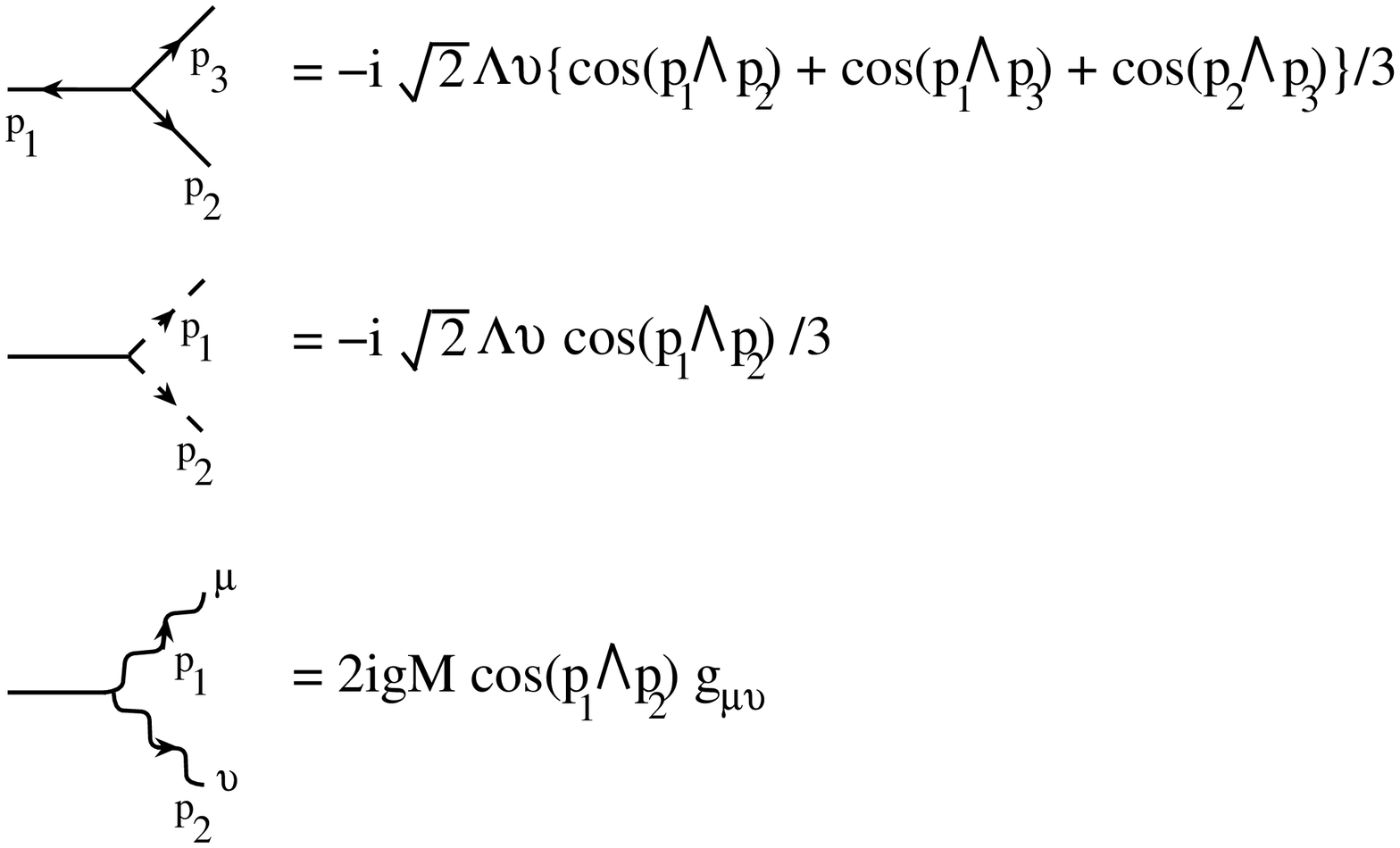,height=8.6cm,width=14.1cm,angle=0}}
\end{figure}

\noindent
\begin{figure}[htbp]
\centerline{
\psfig{figure=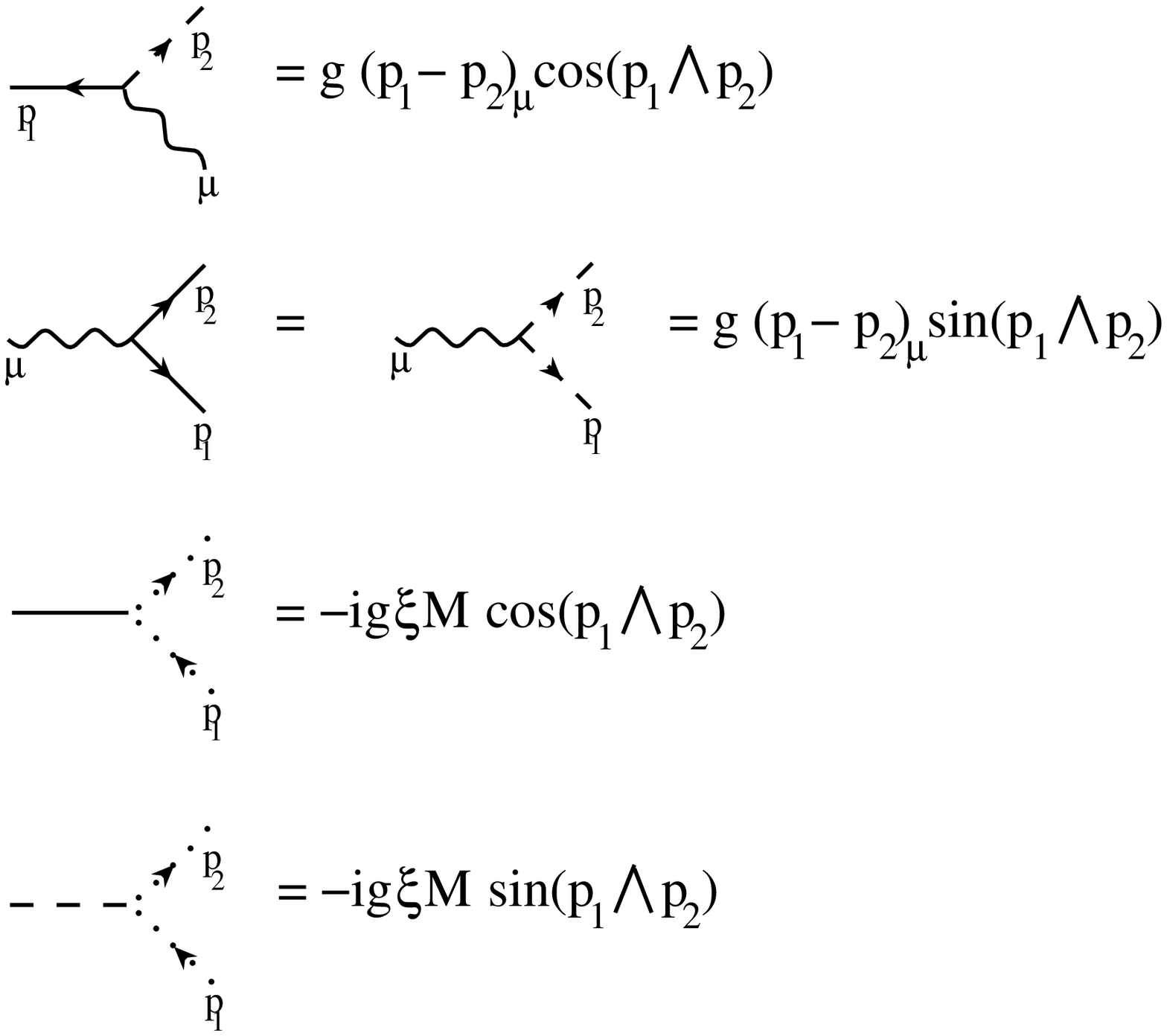,height=9.1cm,width=13.6cm,angle=0}}
\vspace*{-0.1cm}
\centerline{
\psfig{figure=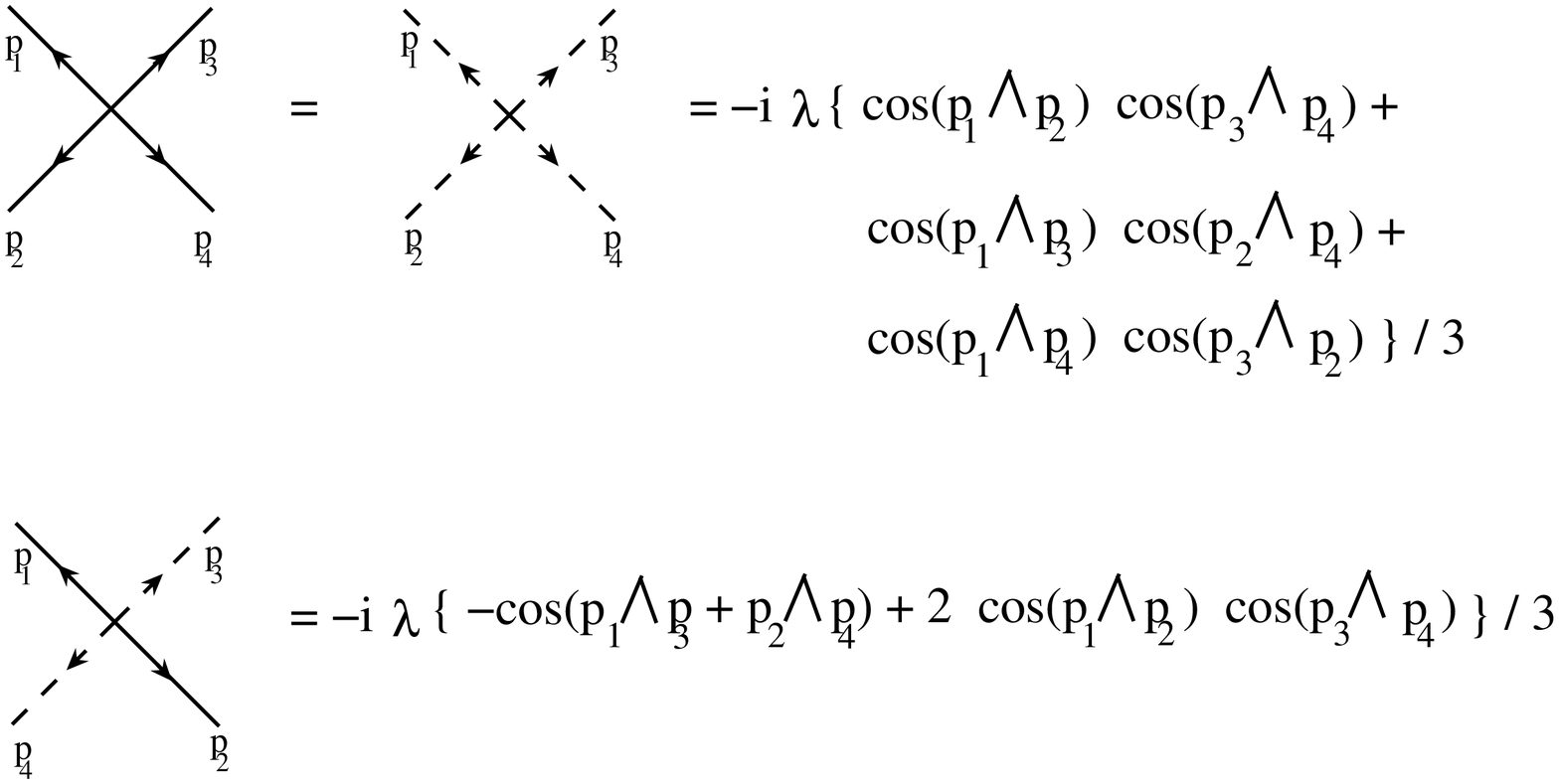,height=8.6cm,width=15.1cm,angle=0}}
\vspace*{-1.5cm}
\centerline{
\psfig{figure=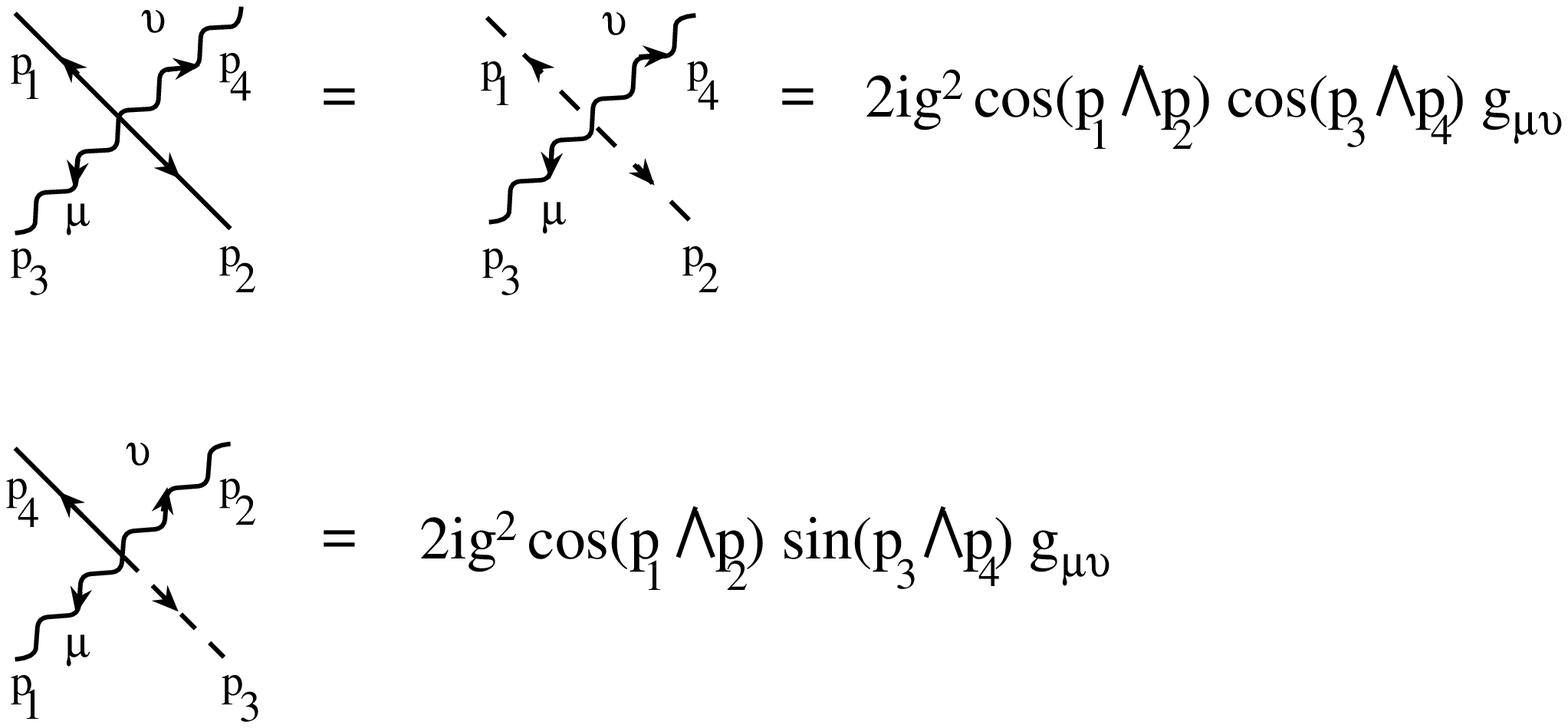,height=8.0cm,width=14.3cm,angle=0}}
\end{figure}

\noindent
\begin{figure}[htbp]
\centerline{
\psfig{figure=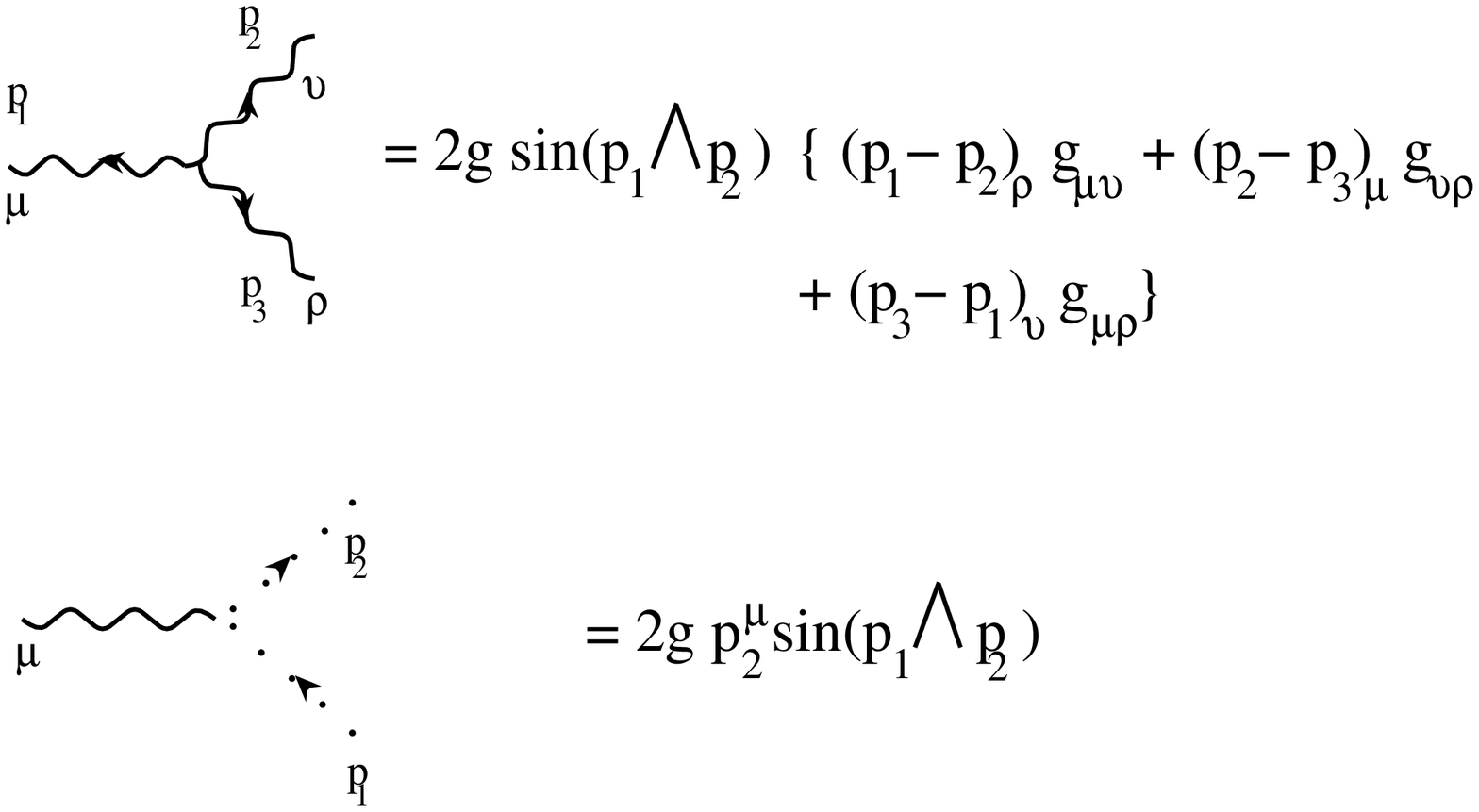,height=7.0cm,width=14.0cm,angle=0}}
\vspace*{-0.2cm}
\centerline{
\psfig{figure=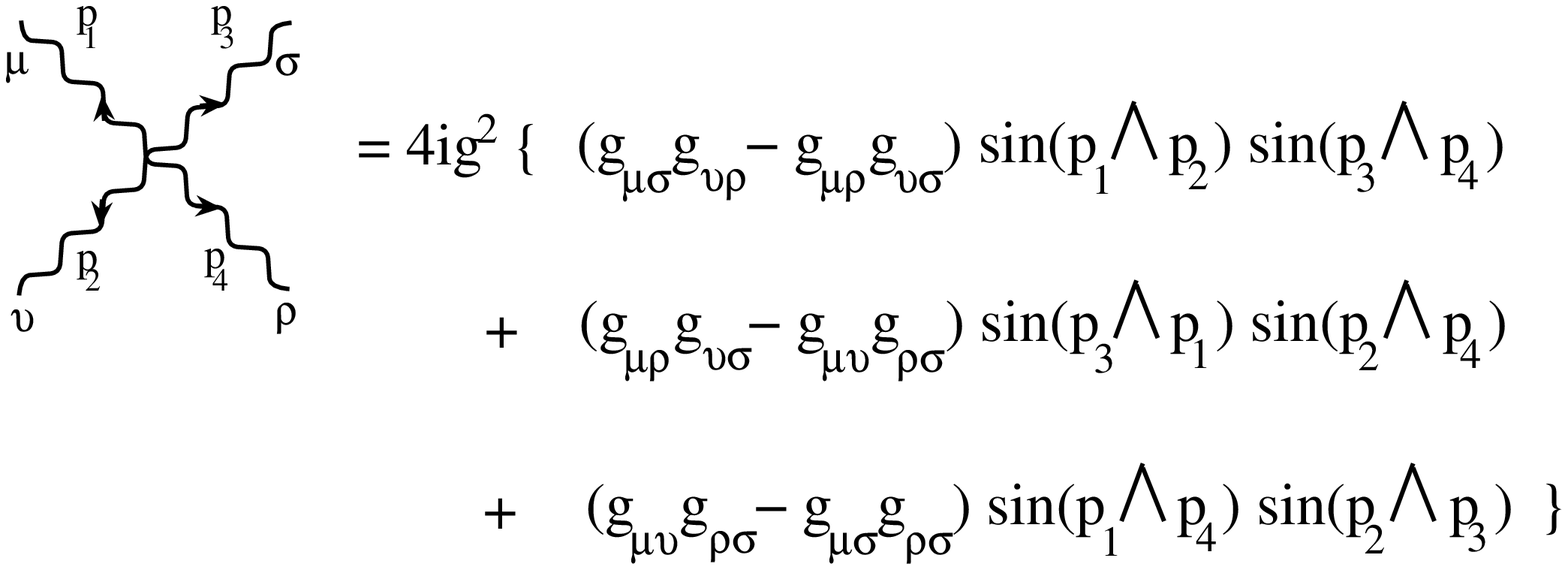,height=6.0cm,width=14.8cm,angle=0}}
\end{figure}

\newpage

\noindent{\Large\bf Appendix B: 1PI diagramatic contributions}

We present here the topologies contributing to the various 1PI divergences.  To save space only the distinct topologies are given; diagrams related to shown diagrams by crossing symmetries are not listed.

\vspace*{-0.7cm}
\noindent
\begin{figure}[htbp]
\centerline{
\psfig{figure=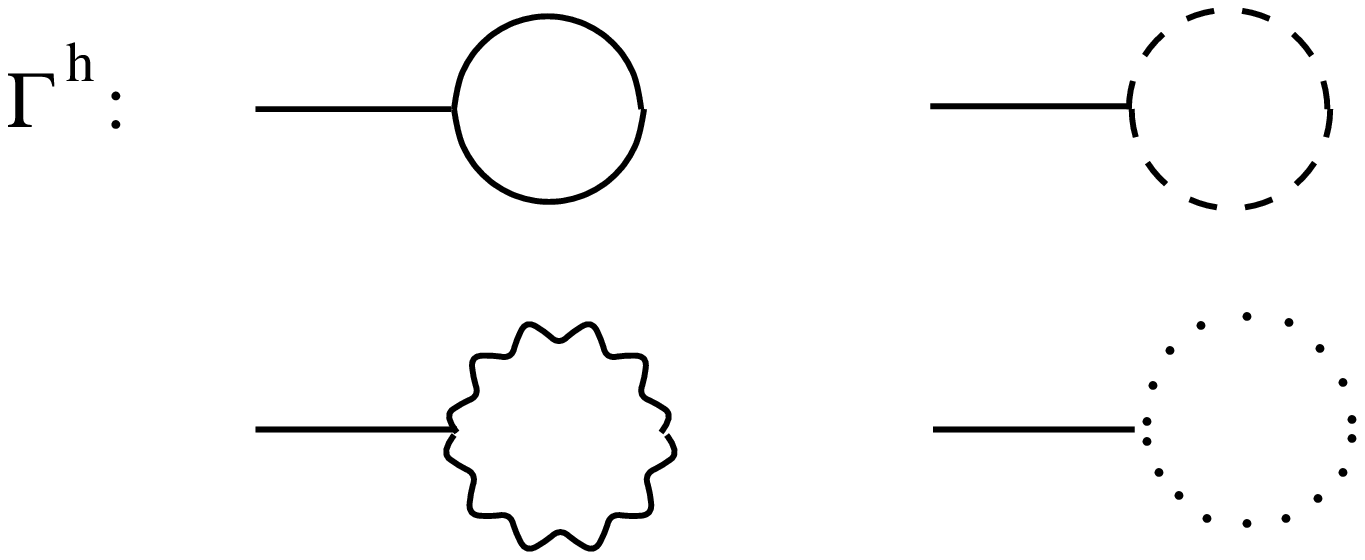 ,height=3.7cm,width=9.0cm,angle=0}}
\vspace*{-0.8cm}
\centerline{
\psfig{figure=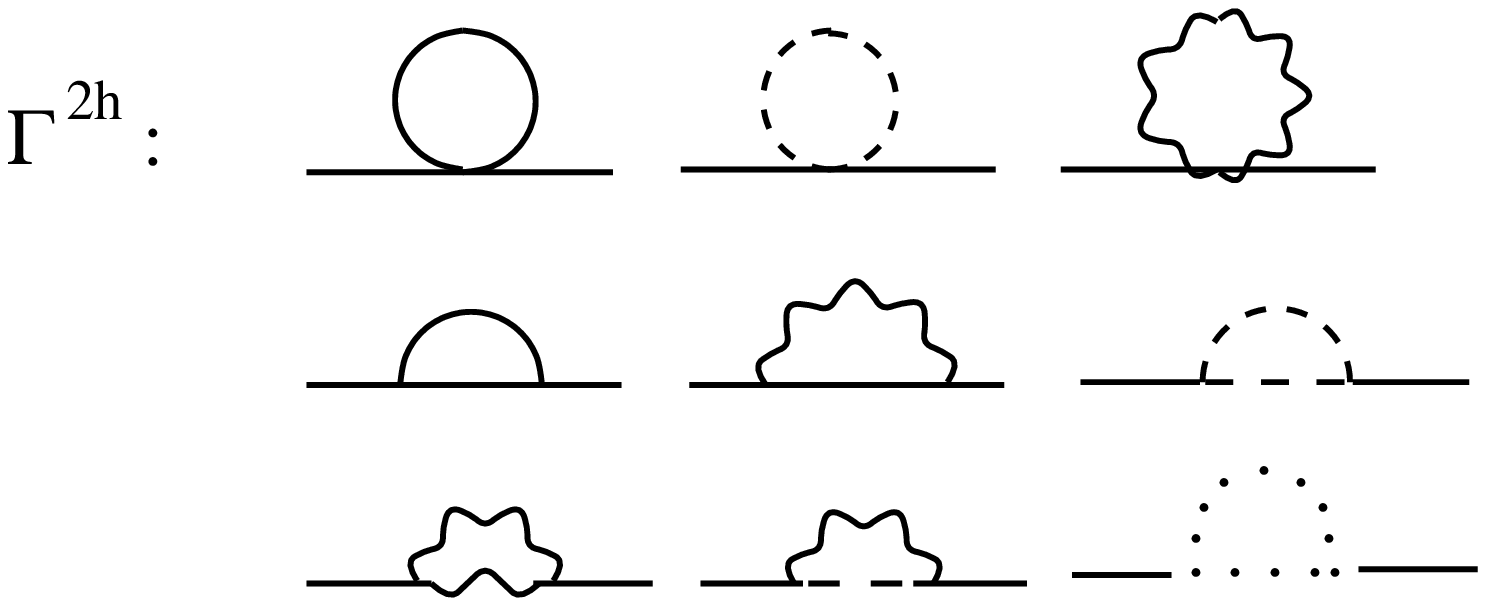 ,height=3.5cm,width=9.5cm,angle=0}}
\vspace*{-0.0cm}
\centerline{
\psfig{figure=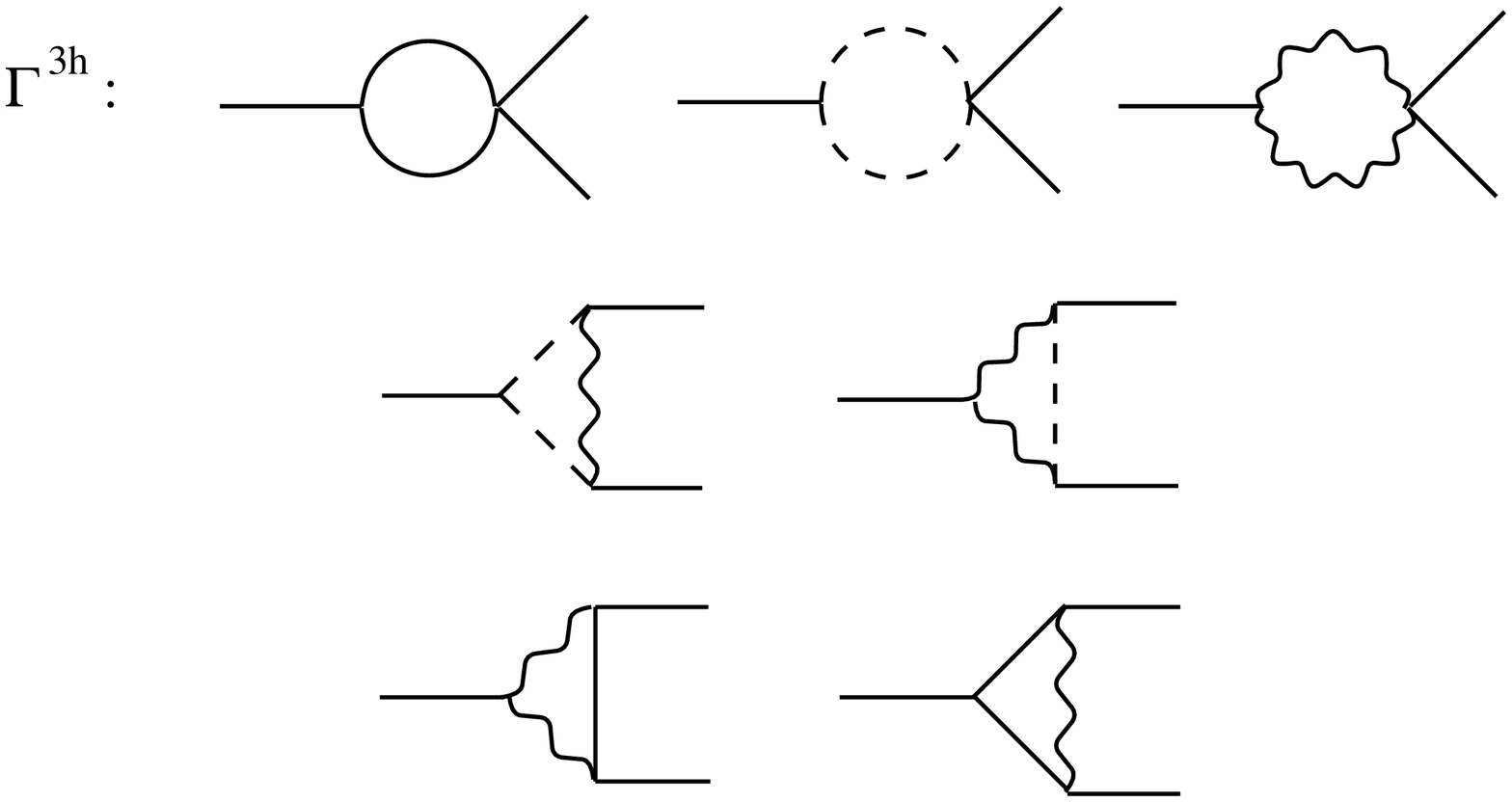 ,height=5.2cm,width=11.0cm,angle=0}}
\vspace*{-0.3cm}
\centerline{
\psfig{figure=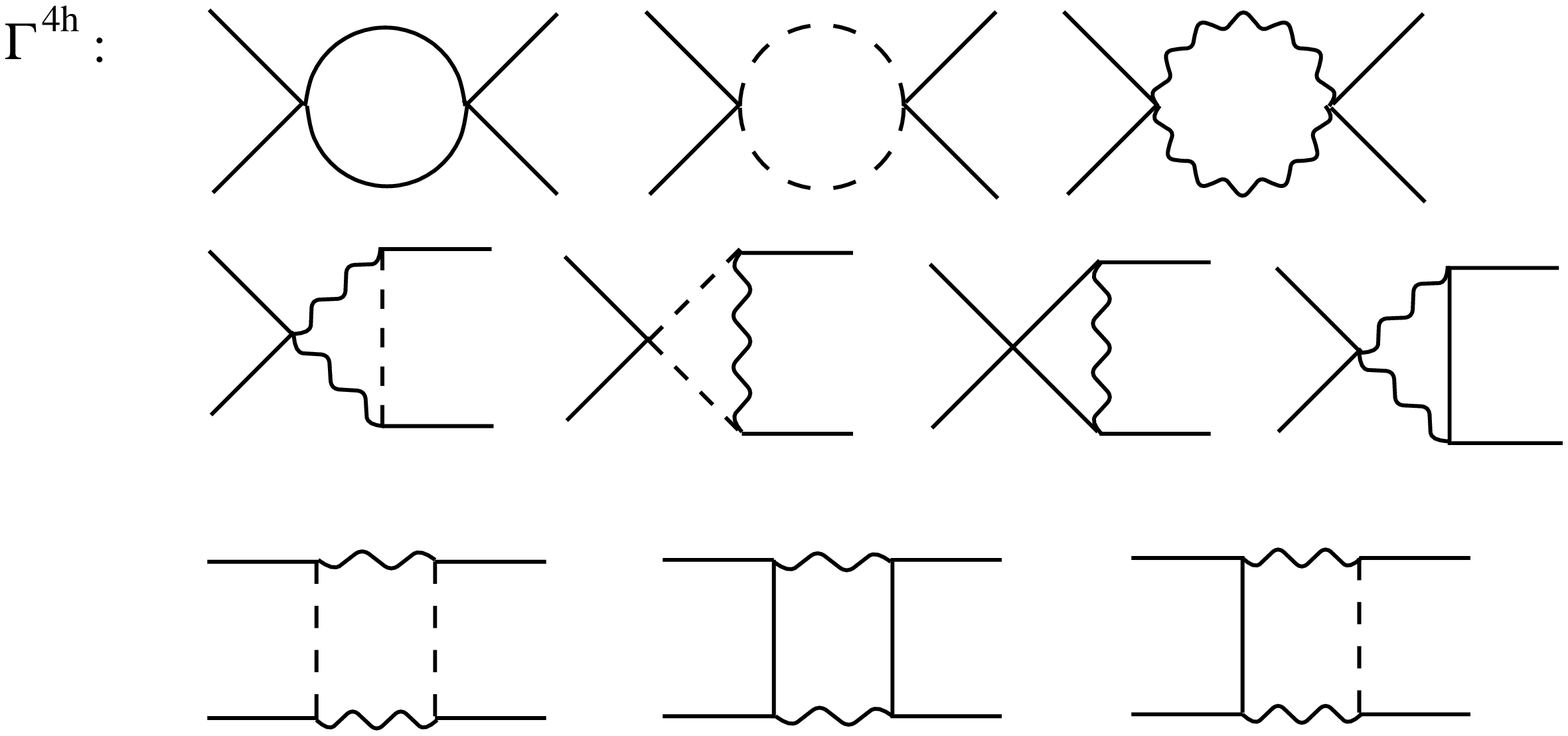 ,height=5.2cm,width=11.0cm,angle=0}}
\end{figure}

\vspace*{-5.8cm}
\noindent
\begin{figure}[htbp]
\vspace*{-0.2cm}
\centerline{
\psfig{figure=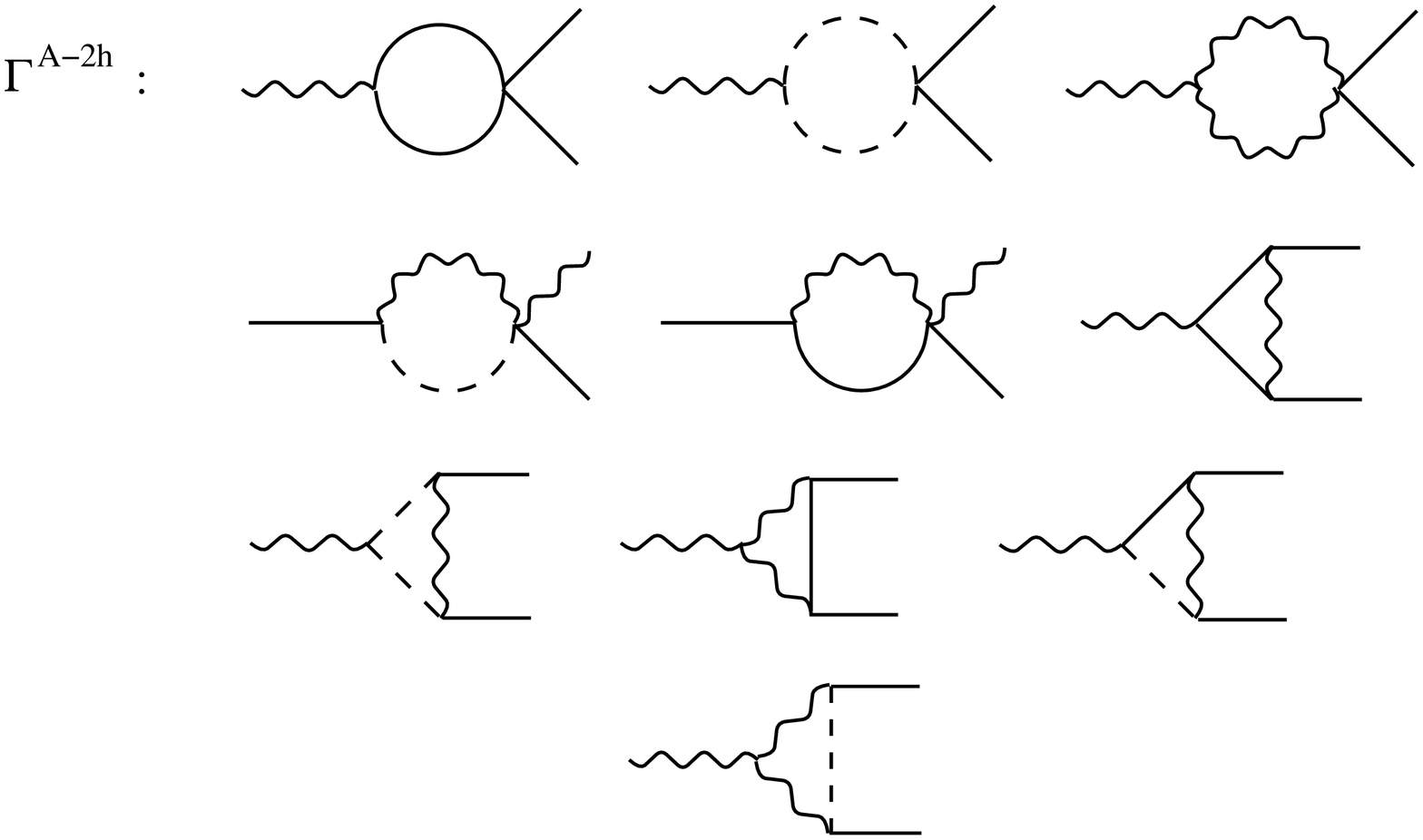 ,height=5.8cm,width=11.5cm,angle=0}}
\vspace*{0.2cm}
\centerline{
\psfig{figure=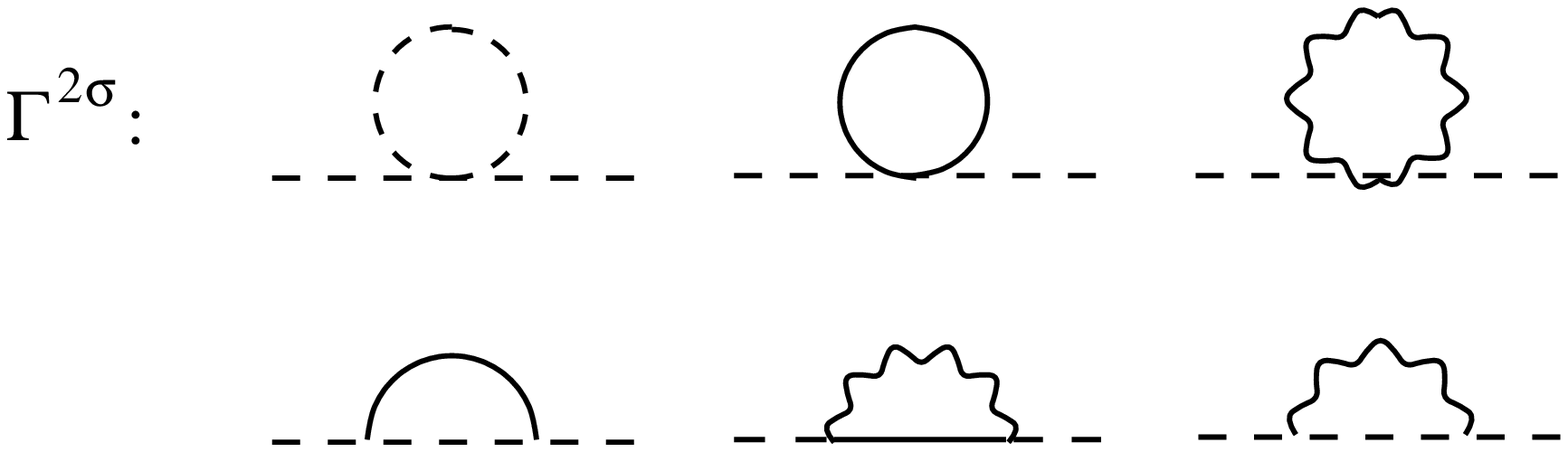 ,height=3.0cm,width=10.5cm,angle=0}}
\vspace*{-0.2cm}
\centerline{
\psfig{figure=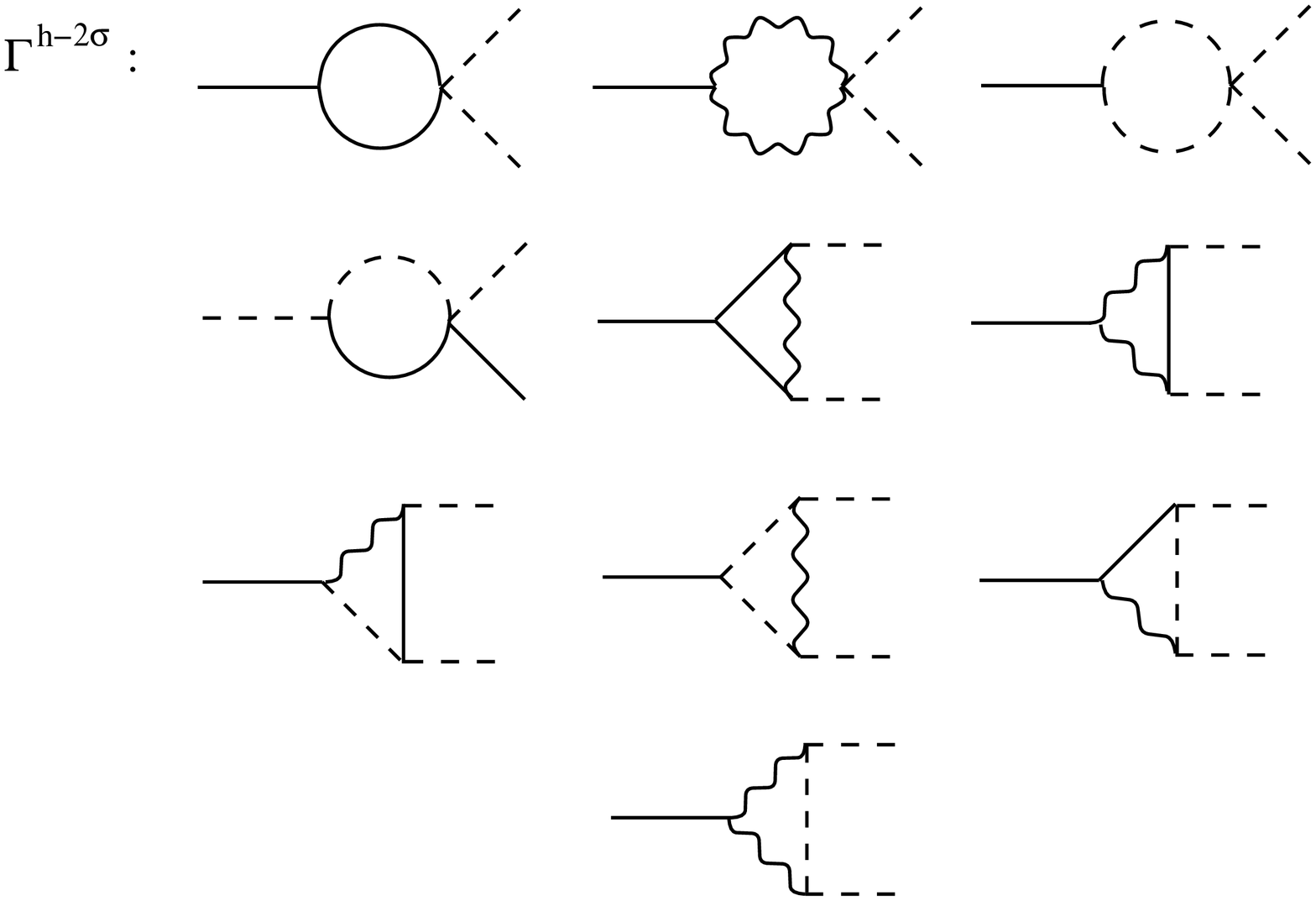 ,height=6.0cm,width=11.0cm,angle=0}}
\vspace*{-0.2cm}
\centerline{
\psfig{figure=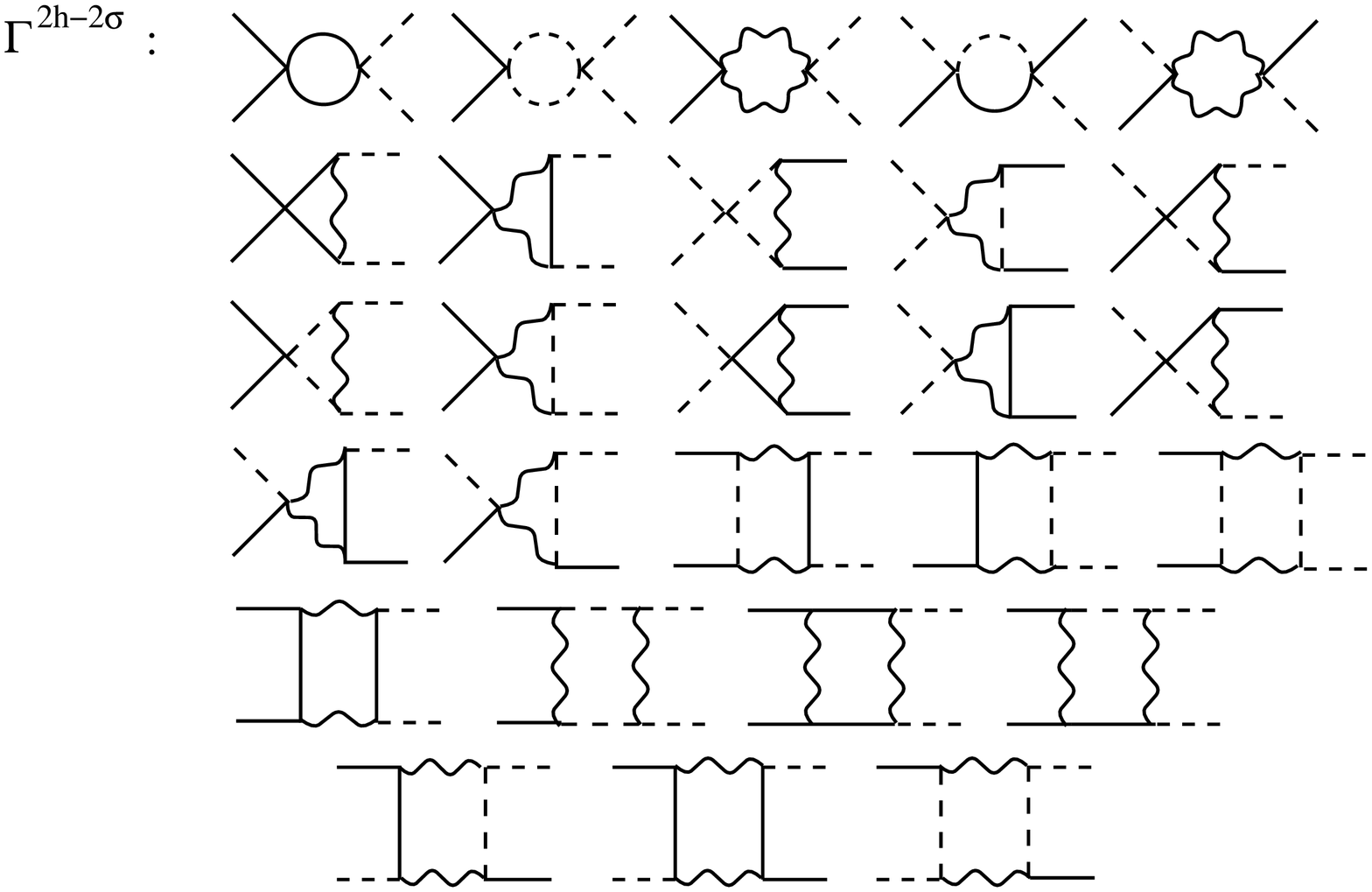 ,height=7.5cm,width=13.0cm,angle=0}}
\end{figure}

\vspace*{-5.8cm}
\noindent
\begin{figure}[htbp]
\vspace*{-0.1cm}
\centerline{
\psfig{figure=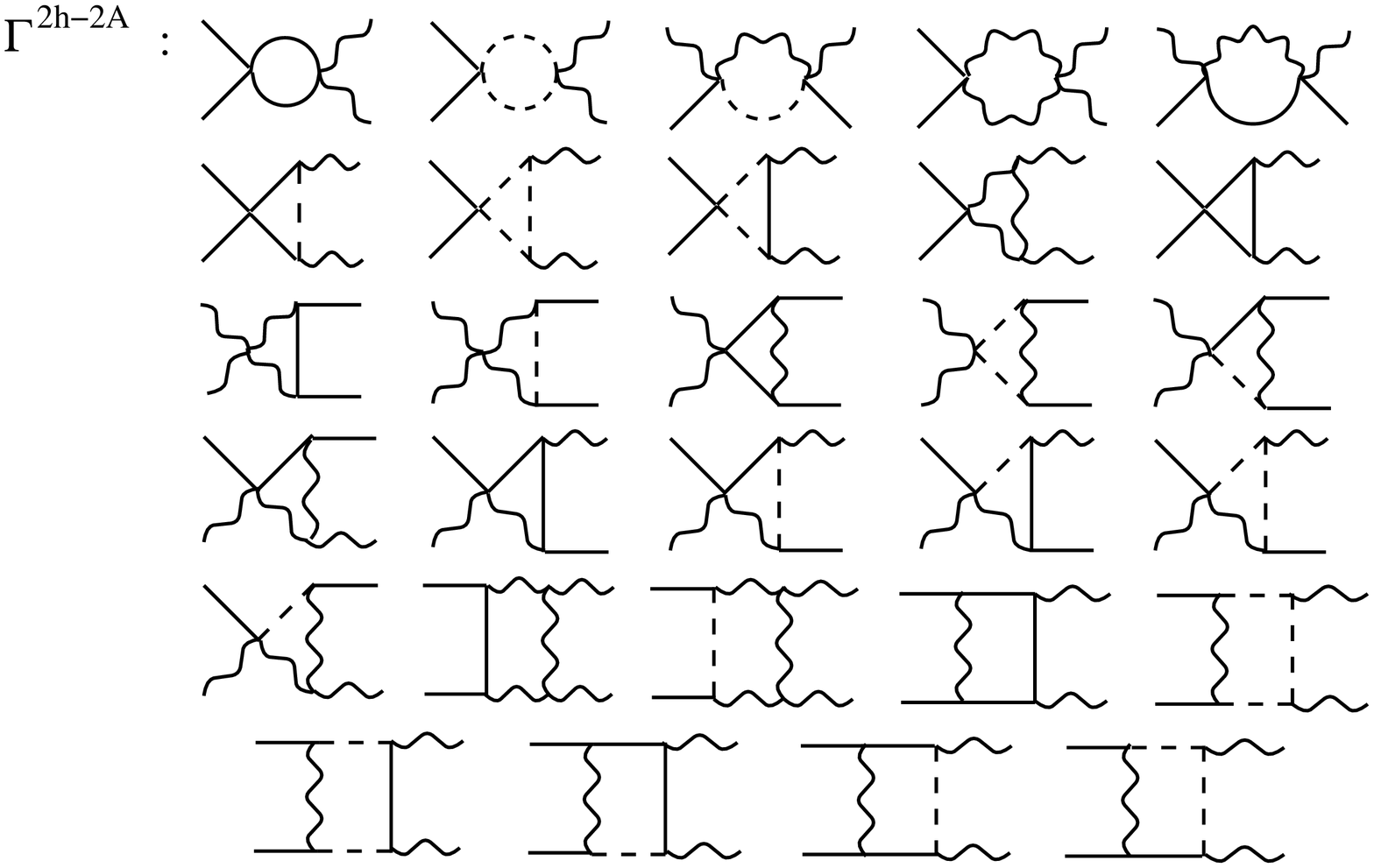 ,height=8.5cm,width=13.5cm,angle=0}}
\vspace*{-0.1cm}
\centerline{
\psfig{figure=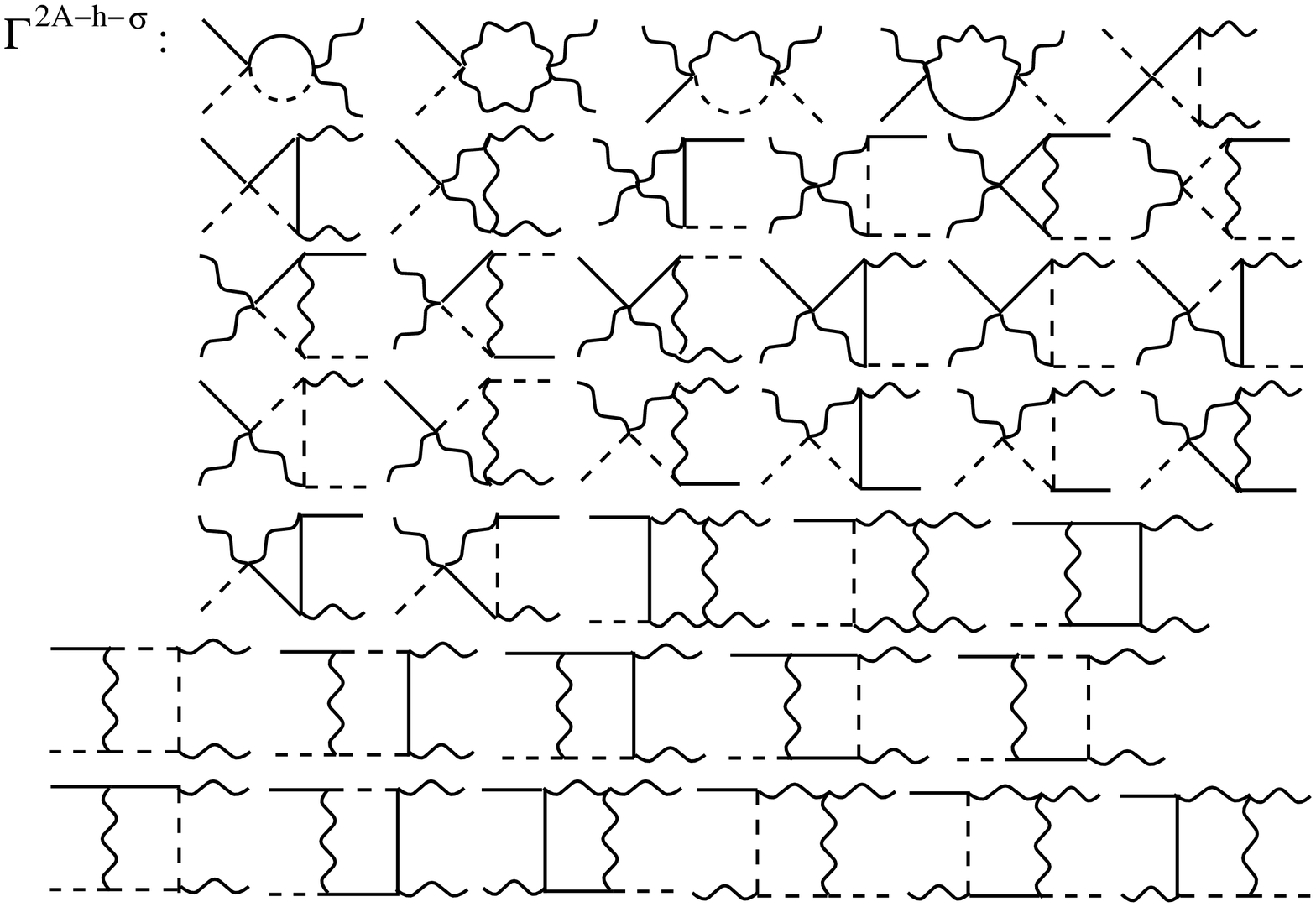 ,height=8.8cm,width=13.5cm,angle=0}}
\end{figure}

\vspace*{-5.8cm}
\noindent
\begin{figure}[htbp]
\vspace*{0.1cm}
\centerline{
\psfig{figure=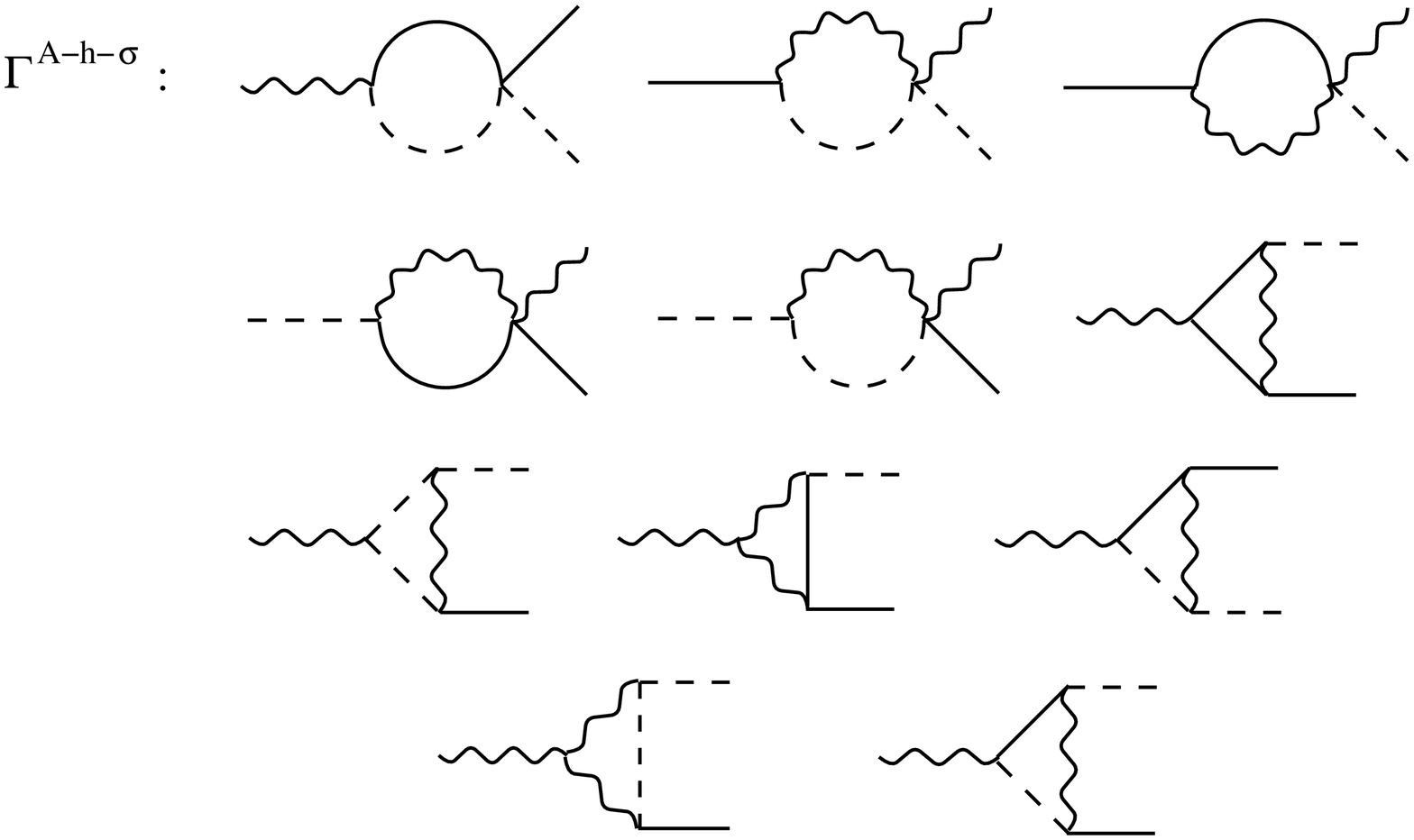 ,height=5.5cm,width=10.0cm,angle=0}}
\vspace*{-0.1cm}
\centerline{
\psfig{figure=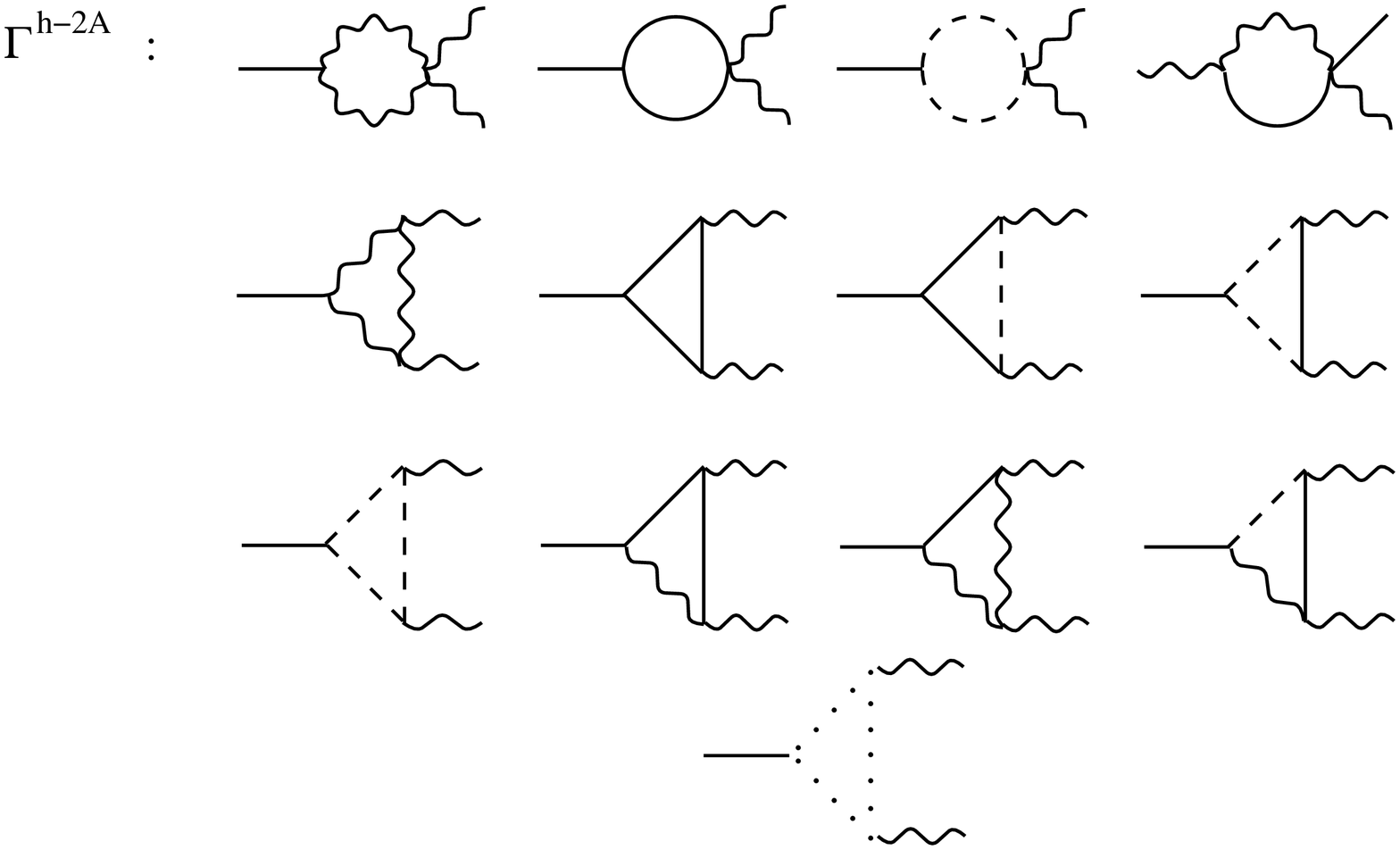 ,height=7.5cm,width=12.5cm,angle=0}}
\vspace*{-0.1cm}
\centerline{
\psfig{figure=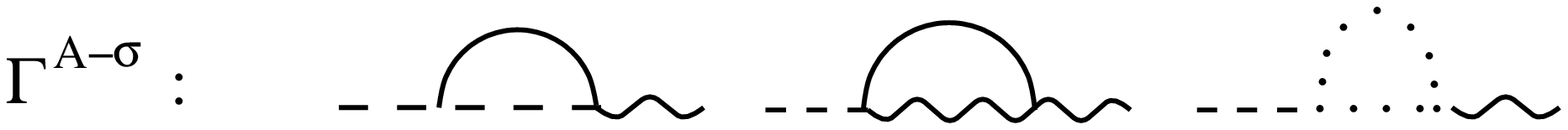 ,height=2.0cm,width=9.0cm,angle=0}}
\vspace*{-0.3cm}
\centerline{
\psfig{figure=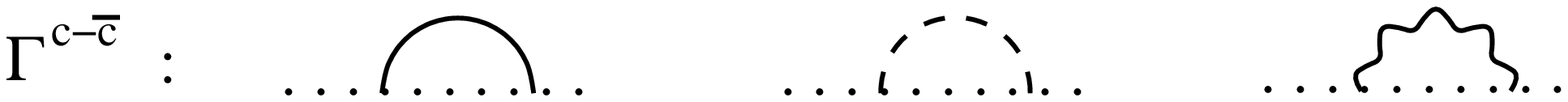 ,height=2.0cm,width=9.5cm,angle=0}}
\vspace*{-0.3cm}
\centerline{
\psfig{figure=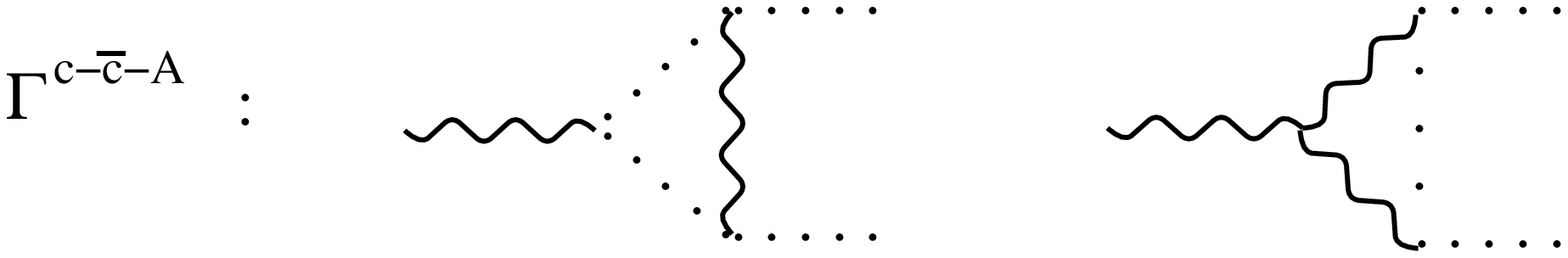 ,height=2.5cm,width=9.5cm,angle=0}}
\end{figure}

\vspace*{-5.8cm}
\noindent
\begin{figure}[htbp]
\vspace*{-0.3cm}
\centerline{
\psfig{figure=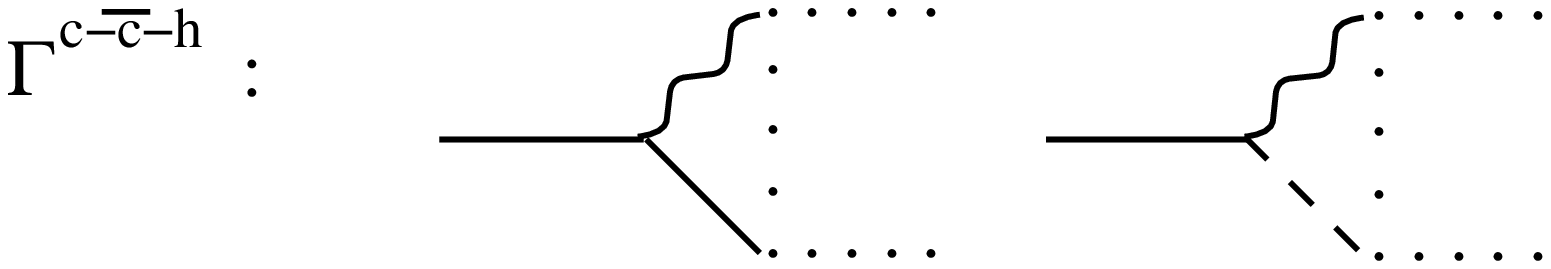 ,height=2.0cm,width=9.0cm,angle=0}}
\vspace*{-0.3cm}
\centerline{
\psfig{figure=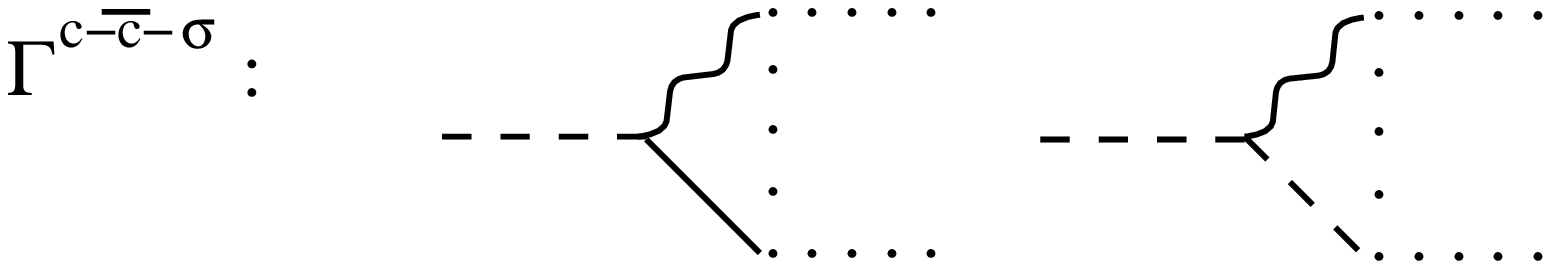 ,height=2.0cm,width=9.0cm,angle=0}}
\vspace*{0.1cm}
\centerline{
\psfig{figure=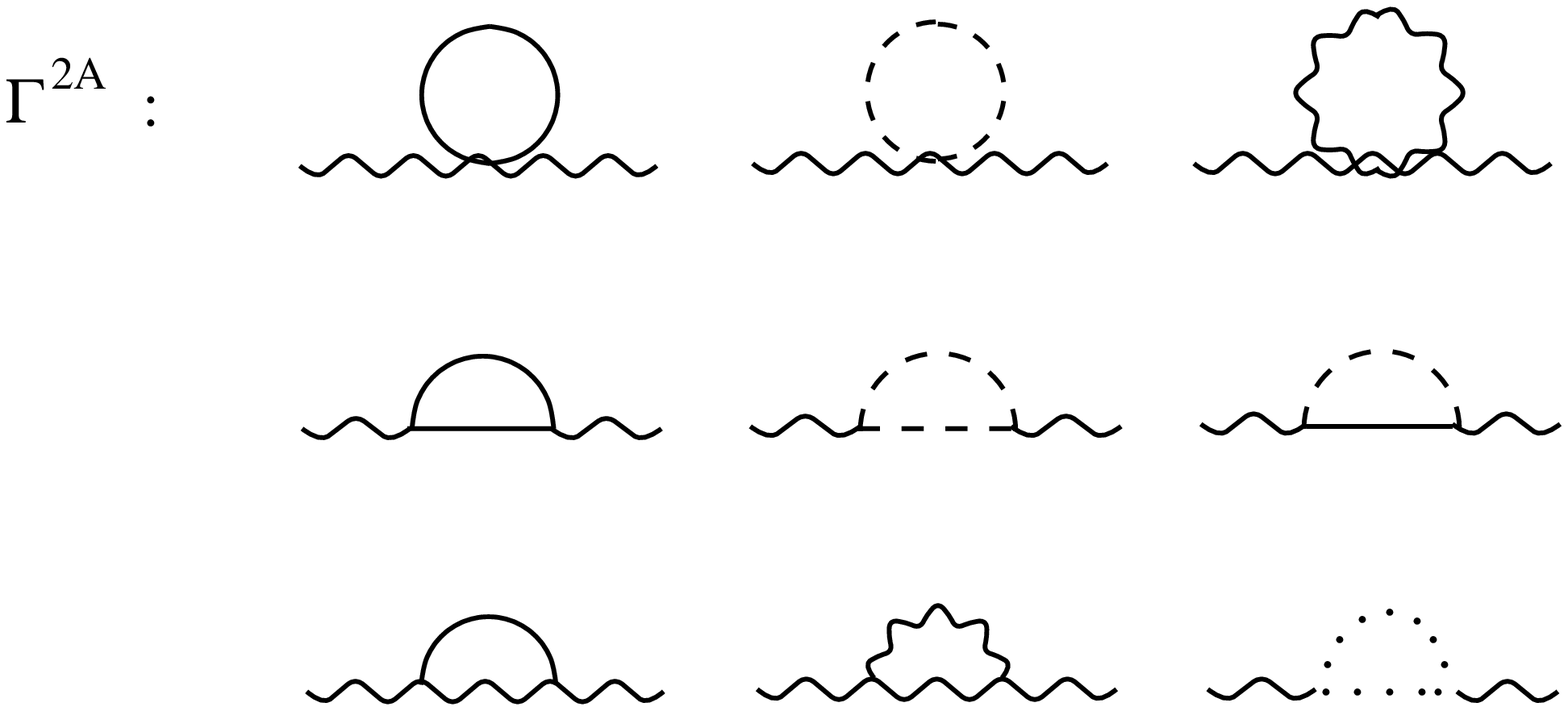 ,height=5.5cm,width=10.0cm,angle=0}}
\vspace*{-0.1cm}
\centerline{
\psfig{figure=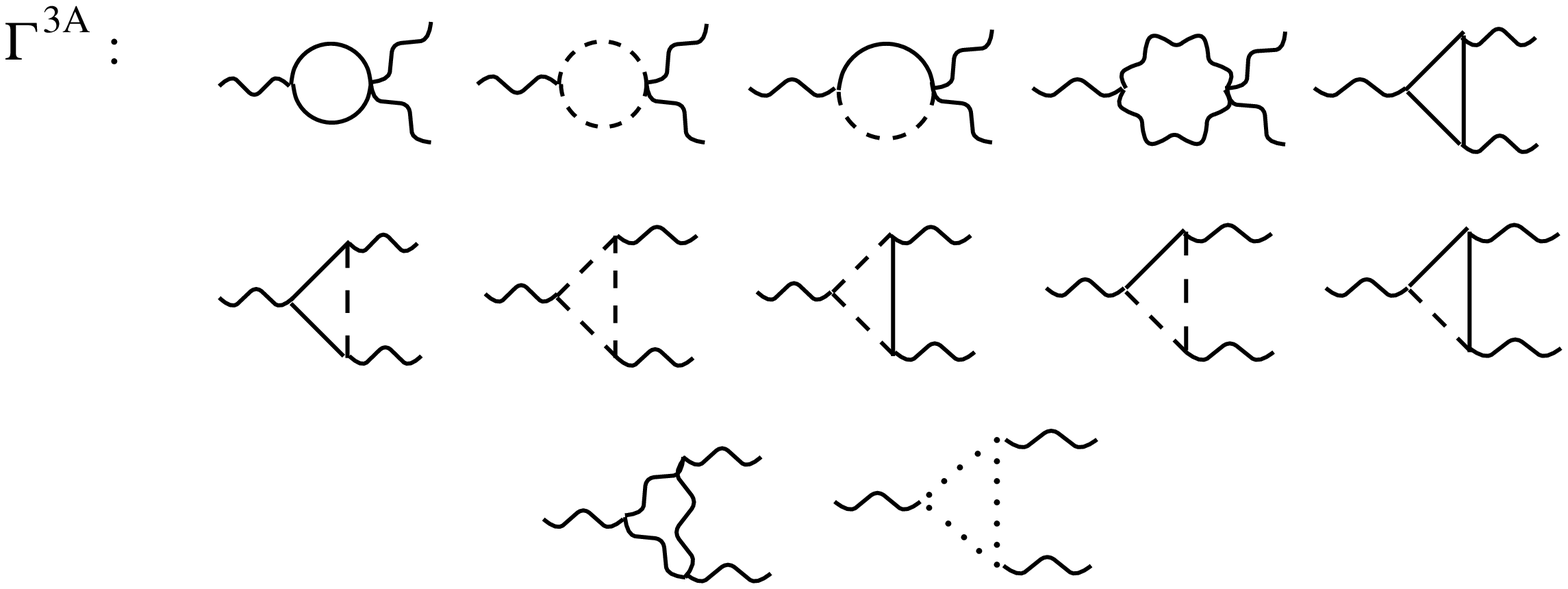 ,height=6.0cm,width=12.0cm,angle=0}}
\end{figure}

\noindent
\begin{figure}[htbp]
\vspace*{-0.1cm}
\centerline{
\psfig{figure=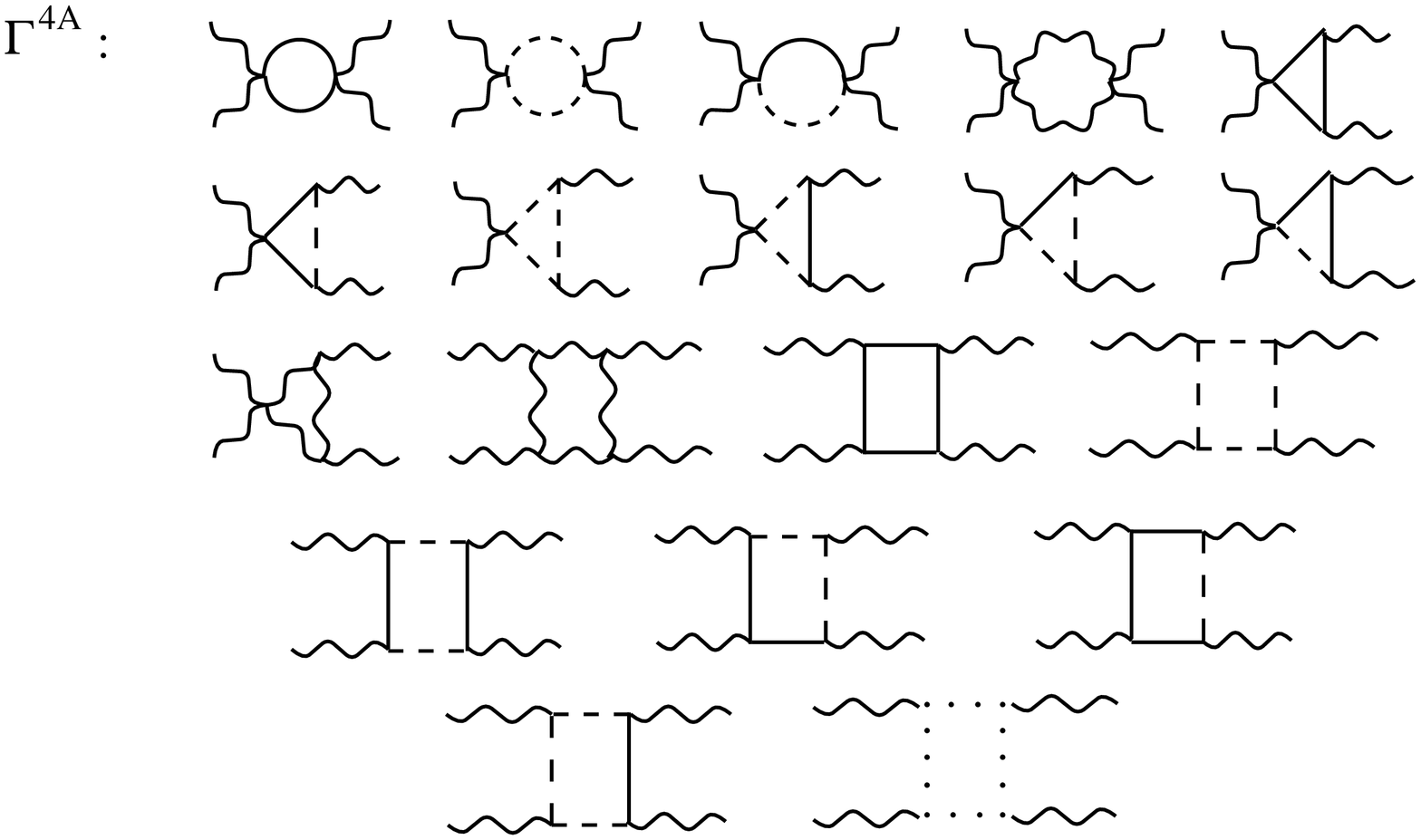 ,height=7.5cm,width=11.0cm,angle=0}}
\vspace*{-0.3cm}
\end{figure}
\end{document}